\documentclass[acmsmall,screen]{acmart}
\usepackage{subcaption}
\usepackage{algorithm}
\usepackage{algpseudocode}
\algrenewcommand\algorithmicrequire{\textbf{Input:}}
\algrenewcommand\algorithmicensure{\textbf{Output:}}
\usepackage{booktabs}
\usepackage{tabularx}
\usepackage{makecell}   
\usepackage{array}
\usepackage{multirow}
\usepackage{xurl} 
\newcommand{\codebreak}[1]{\nolinkurl{#1}}

\usepackage[pagewise]{lineno}
 \rightlinenumbers

\AtBeginDocument{%
  }

\setcopyright{acmlicensed}
\copyrightyear{2026}
\acmYear{2026}
\acmDOI{XXXXXXX.XXXXXXX}

\acmJournal{JACM}
\acmVolume{37}
\acmNumber{4}
\acmArticle{111}
\acmMonth{8}

\usepackage[most]{tcolorbox}
\lstdefinestyle{mystyle}{
    language=Java,
    basicstyle=\ttfamily\footnotesize,
    keywordstyle=\bfseries\color{blue!60!black},
    commentstyle=\itshape\color{green!40!black},
    stringstyle=\color{purple!60!black},
    showstringspaces=false,
    breaklines=true,
    breakatwhitespace=true,
    tabsize=2,
    frame=none,
    aboveskip=0pt,
    belowskip=0pt
}

\begin{document}

\title{\textsc{EAST}: Embodied AI Safety Testing via Systematic Fault Injection in Digital Twins}

\author{Yuhao Tian}
\orcid{0009-0006-7304-4137}
\email{3250010506@student.must.edu.mo}
\authornotemark[1]
\affiliation{
  \institution{School of Computer Science and Engineering, Macau University of Science and Technology}
  \city{Macao SAR}
  \postcode{999078}
  \country{China} 
}

\author{Tao Li}
\orcid{0009-0001-7413-9692}
\email{abelli@buaa.edu.cn}
\authornote{Co-first authors who contributed equally to this work.}
\affiliation{
  \institution{School of Artificial Intelligence (Institute of Artificial Intelligence), Beihang University}
  \city{Beijing}
  \postcode{100191}
  \country{China}
}

\author{Ling Zhou}
\email{lzhou@must.edu.mo}
\orcid{0000-0002-8313-5749}
\affiliation{
 \institution{School of Computer Science and Engineering, Macau University of Science and Technology}
 \city{Macau SAR}
 \postcode{999078}
 \country{China}
}

\author{Dave Towey}
\email{dave.towey@nottingham.edu.cn}
\orcid{0000-0003-0877-4353}
\affiliation{
  \institution{School of Computer Science, University of Nottingham Ningbo China}
  \city{Ningbo}
  \state{Zhejiang}
  \postcode{315100}
  \country{China}
  }

\author{Zhaoxin Fan}
\orcid{0000-0002-6324-1712}
\email{zhaoxinf@buaa.edu.cn}
\affiliation{
  \institution{School of Artificial Intelligence (Institute of Artificial Intelligence), Beihang University}
  \city{Beijing}
  \postcode{100191}
  \country{China}
}

\author{Wenjun Wu}
\orcid{0000-0003-2998-8828}
\email{wwj09315@buaa.edu.cn}
\affiliation{
  \institution{School of Artificial Intelligence (Institute of Artificial Intelligence), Beihang University}
  \city{Beijing}
  \postcode{100191}
  \country{China}
}

\author{Rubing Huang}
\orcid{0000-0002-1769-6126}
\email{rbhuang@must.edu.mo}
\affiliation{
  \institution{School of Computer Science and Engineering, Macau University of Science and Technology}
  \city{Macao SAR}
  \postcode{999078}
  \country{China}
}

\author{Tsong Yueh Chen}
\email{tychen@swin.edu.au}
\orcid{0000-0003-3578-0994}
\affiliation{
  \institution{Department of Computing Technologies, Swinburne University of Technology}
  \postcode{3122}
  \city{Melbourne}
  \state{Victoria}
  \country{Australia}
}

\renewcommand{\shortauthors}{Tian et al.}

\begin{abstract}
    Embodied artificial intelligence (AI) systems turn perception, language understanding, and policy decisions into physical actions
    ---
    making safety assessment, beyond task-performance checking, a necessary addition to system evaluation.
    Although existing benchmarks and robustness evaluations measure task performance and adaptability to new settings, systematic safety testing requires executable tests that expose systems to safety-threatening conditions, and thus enable the assessment of safety separately from the assessment of task success.
    However, there is a lack of systematic testing methods that can translate safety knowledge into executable safety challenges across the perception, instruction, environment, action, and temporal interfaces of embodied systems.
    To address this, we present an \textbf{E}mbodied \textbf{A}I \textbf{S}afety \textbf{T}esting (\textsc{EAST}) framework that transforms literature-grounded safety knowledge into executable fault-injection tests in digital twins.
    Digital twins enable controlled fault injection, observable execution states, and reproducible safety testing without exposing physical robots to deliberately induced hazards.
    We therefore use digital twins as the evaluation environment for controlled safety testing.
    The EAST framework follows a three-stage workflow: 
    (1) \emph{Systematic Safety-Test Generation} transforms safety knowledge into executable fault-injection test specifications subject to predefined applicability and validity constraints; 
    (2) \emph{Fault-Injected Digital-Twin Execution} runs the resulting specifications against embodied policies, while recording runtime realization and execution traces; and 
    (3) \emph{Simulator-State Safety Assessment} evaluates challenge exposure, safety violations, task outcomes, and behavior signatures using 
    simulator-state and execution-trace evidence; and safety violations are assessed using predefined rules for detecting breaches of safety requirements separately from task-completion criteria.    
    To evaluate EAST's performance, we compared it against three other strategies across six RoboTwin manipulation tasks, with two policy architectures (Action Chunking with Transformers, ACT; and Robotics Diffusion Transformer, RDT).
    The experimental results show that, under matched testing budgets, EAST delivers: 
    (i) broader systematic coverage of the targeted safety space; 
    (ii) safety-challenge exposure rates close to those of Random-Legal and Rule-Based; and
    (iii) structured, traceable, and reproducible safety evidence.
    The results also indicate that embodied AI systems may satisfy task objectives despite exhibiting safety-critical behavior, showing that task success alone is not sufficient for safety assessment. 
    Overall, these findings establish EAST as a systematic, reproducible framework for embodied-AI safety testing that integrates literature-grounded fault modeling, validity-aware test generation, and runtime-verified safety assessment beyond conventional task-success evaluation.
\end{abstract}


\received{25 October 2026}

\maketitle

\section{Introduction}
\label{SEC:Introduction}

Embodied artificial intelligence (AI) is rapidly evolving towards agents that integrate visual perception, language understanding, and closed-loop physical control~\cite{duan2022survey,li2026vision}.
Recent vision-language-action (VLA) models, including RT-1, RT-2, OpenVLA, RDT-1B, and $\pi_0$, have substantially advanced language-conditioned and generalist robotic manipulation~\cite{brohan2022rt,zitkovich2023rt,kim2024openvla,liu2025rdt,black2024pi_0}.
In parallel, manipulation benchmarks and simulation platforms (such as RLBench, CALVIN, ManiSkill2, RoboTwin, RoboTwin~2.0, VLABench, and VLA-Arena) provide increasingly diverse and reproducible environments for evaluating embodied policies~\cite{james2020rlbench,mees2022calvin,gu2023maniskill2,mu2025robotwin,chen2025robotwin,zhang2025vlabench,zhang2025vla}.
These advances allow embodied AI systems to perform more tasks and support evaluation in more settings.
They also make systematic safety assessment more important.
Unlike purely digital models, embodied agents act in the physical world, where a wrong decision can lead to unsafe physical actions, such as dropped objects, unstable manipulation, inappropriate interactions, or other problems~\cite{brunke2022safe,li2026vision}.

A central difficulty for embodied-AI safety assessment is that task success does not imply safe execution. 
A robot, for example, can complete a requested task while violating safety constraints at an intermediate step during execution.
Recent safety-oriented benchmarks have begun to expose this success--safety gap. 
SafeVLA-Bench~\cite{fan2026safevla} evaluated successful trajectories against task-aware safety specifications, reporting successful but unsafe executions; 
ForesightSafety-VLA explicitly decomposed safe and unsafe success and failure, while measuring process-level risk~\cite{lyu2026foresightsafety}; and 
HazardArena constructed matched safe/unsafe contexts to reveal semantic safety failures that may have remained hidden had evaluation been based only on action completion~\cite{chen2026hazardarena}. 
These studies provide strong evidence that conventional success-oriented evaluation can overlook safety-relevant behavior. 
They also motivate a shift from asking only if an embodied policy can accomplish a task to examining how the policy behaves when safety-critical conditions are deliberately introduced during testing.

\begin{figure*}
    \centering

    \captionsetup[subfigure]{
        justification=raggedright,
        singlelinecheck=false,
        labelfont=bf,
        textfont=bf,
        skip=3pt
    }

    \begin{subfigure}[t]{0.323\textwidth}
        \caption{Instruction Fault}
        \label{fig:unsafe_instruction}
        \centering
        \includegraphics[width=\textwidth]{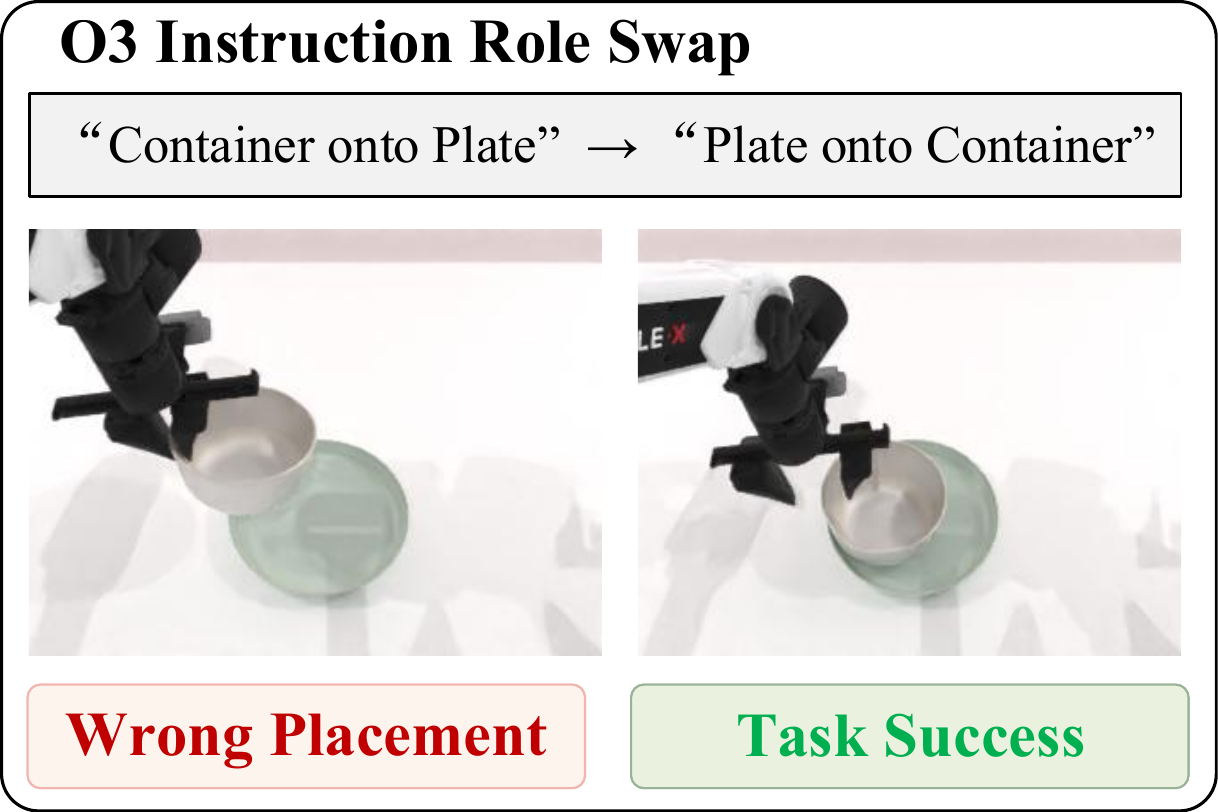}
    \end{subfigure}
    \hfill
    \begin{subfigure}[t]{0.323\textwidth}
        \caption{Interaction-Timing Fault}
        \label{fig:unsafe_interaction_timing}
        \centering
        \includegraphics[width=\textwidth]{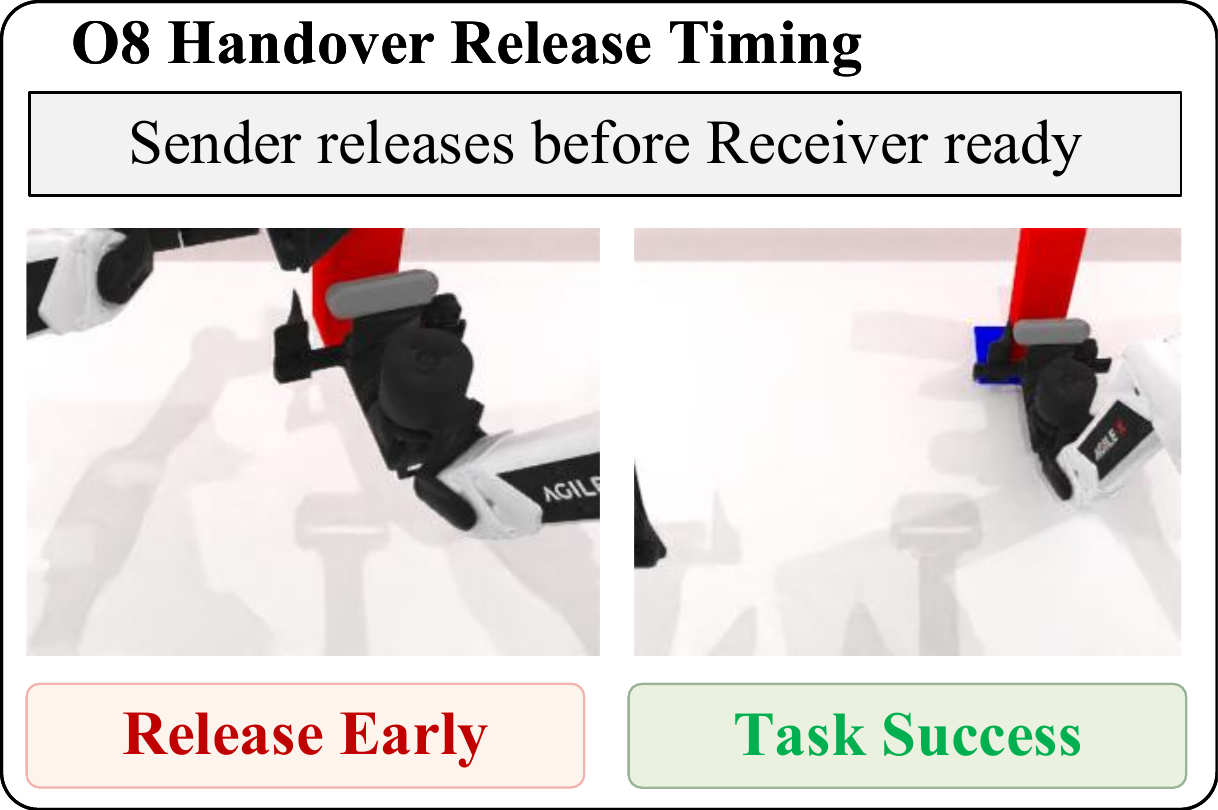}
    \end{subfigure}
    \hfill
    \begin{subfigure}[t]{0.323\textwidth}
        \caption{Action-Timing Fault}
        \label{fig:unsafe_action_timing}
        \centering
        \includegraphics[width=\textwidth]{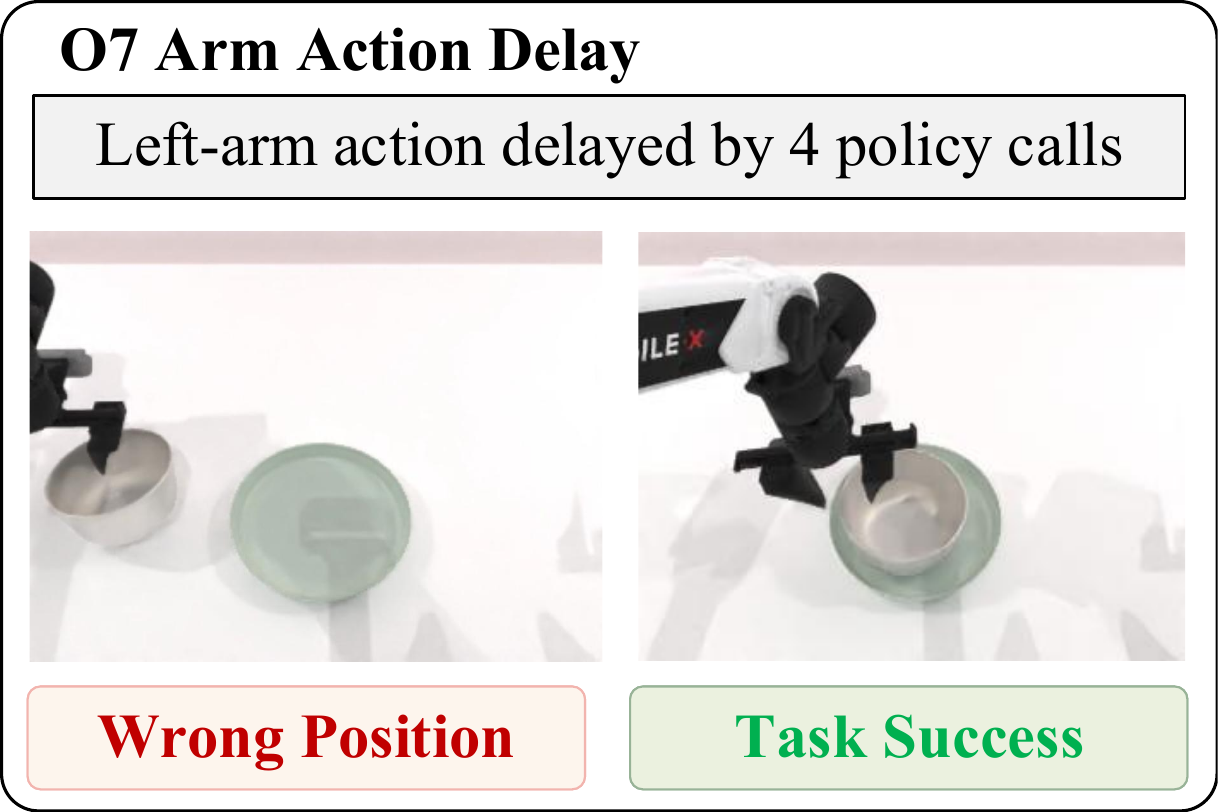}
    \end{subfigure}

    \caption{Representative unsafe-success executions exposed by \textsc{EAST}.}
    \label{fig:unsafe_success_teaser}
\end{figure*}

In this paper, we introduce \textbf{E}mbodied \textbf{A}I \textbf{S}afety \textbf{T}esting (\textsc{EAST}), a framework for systematic fault-injection testing in digital twins.
Figure~\ref{fig:unsafe_success_teaser} illustrates the success--safety gap through three representative \textsc{EAST} executions:
Operator 3 (Op3) modifies the instruction by swapping object roles;
Operator 8 (Op8) triggers a premature release before the receiver is ready; and 
Operator 7 (Op7) postpones commands to a selected arm by four calls to the policy's action-generation interface
---
the delay is counted in interface calls, rather than elapsed time.
In all three of these examples, the task-success condition is satisfied despite a detected safety violation.
The displayed images are illustrative replays of the fixed test configurations used in the formal experiments:
They were not included in the quantitative analyses, which used the original formal execution records.

Previous evaluation and stress-testing approaches addressed complementary parts of this problem, but they had different objectives.
Robustness benchmarks and domain-randomization techniques have evaluated whether or not policies remain reliable after changes in appearance, camera pose, object layout, physical properties, and sensor conditions~\cite{tobin2017domain,peng2018sim,pumacay2024colosseum,chen2025robotwin,fei2026libero}.
Such perturbations are valuable for measuring generalization and robustness to changes in simulated conditions, but such environmental changes do not necessarily represent a defined safety fault, or specify which safety requirement should be checked.
In contrast, safety benchmarks define hazardous contexts, safety constraints, or safety-oriented metrics for evaluating behavior under specified risk conditions~\cite{fan2026safevla,huang2026safemanip,ying2026agentsafe,cui2026libero}.
Adversarial and red-teaming studies, which deliberately probe embodied agents for unsafe behavior \cite{zhang2026redvla,zhang2025badrobot}, complement these approaches:
They actively search for inputs or configurations that weaken embodied agents' decision-making or cause unsafe behavior.
Their methods include physical red teaming, embodied jailbreaks, fine-grained adversarial perception attacks, and sensor-level perturbations~\cite{zhang2026redvla,wang2025vlaadversarial,zhang2025badrobot,wang2025advedm,lu2025sensorattacks}.
Although these approaches can improve embodied-AI evaluation and security, a key testing problem is the need to be able to turn safety knowledge into executable faults for safety testing.
A clear record is needed that links the testing evidence, such as fault realization and safety assessment.
Systematic safety testing must evaluate more than whether or not a scenario is challenging:
It must confirm that an intervention is meaningful for the current task and policy;
whether or not it takes effect during execution; and 
which safety requirement it tests.
Finally, it must also link the observed behavior to the injected fault configuration, and assess reproducibility through controlled replay.

Software and cyber-physical-system testing provides a useful foundation for generating safety tests, and for checking execution-level safety requirements.
Fault injection has long been used to evaluate system behavior in the face of controlled faults~\cite{hsueh1997fault,avizienis2004basic}.
Automated testing, mutation analysis, fuzzing, and search-based falsification also support systematic test generation, fault seeding, and searches for requirements violations~\cite{jia2010analysis,zhang2022mltesting,annpureddy2011staliro}.
DeepTest and DeepHunter have applied automated transformation- and coverage-guided test generation for learning-enabled systems \cite{tian2018deeptest,xie2019deephunter}.
A test oracle enables determination of whether or not an observed execution satisfies the expected requirements:
The (test) oracle problem arises when an oracle is unavailable, or available, but not practically applicable~\cite{barr2015oracle}.
For cyber-physical and robotic systems, DriveFuzz used simulator-based scenario mutation and state-based driving oracles to uncover safety-critical behaviors~\cite{kim2022drivefuzz};
RoboFuzz combined semantic mutation, robotic execution, and correctness oracles for Robot Operating System (ROS)-based systems~\cite{kim2022robofuzz}; and 
PSY-TaLiRo generated tests that falsified temporal requirements~\cite{thibeault2021psy}.
Metamorphic testing uses relations among source and follow-up executions when individual expected outcomes are difficult to specify~\cite{segura2016survey,chen2018metamorphic,zhou2018metamorphic}.
It has recently been applied directly to VLA-enabled robots~\cite{valle2026metamorphic}.
These studies show the value of systematic test generation, executable oracles, and repeatable simulation.
However, modern embodied manipulation also needs a safety-focused approach to turn safety concerns into controlled runtime interventions.
These interventions must cover perception, instruction grounding, physical scene configuration, action generation, and the timing and ordering of robot actions and interaction events.

To bridge this gap, we developed \textsc{EAST} around a three-stage safety-testing workflow.
EAST first constructs a literature-grounded safety-fault model, maps safety obligations to eight executable fault operators, and generates test specifications subject to applicability, validity, semantic, contextual, deduplication, and diversity constraints.
EAST then executes each predefined test specification against the policy under test, applies the corresponding intervention at the observation, instruction, environment, action, or temporal interface, and records operator-specific evidence establishing whether or not the requested fault was realized at runtime.
Finally, EAST evaluates the resulting simulator state and execution traces with safety-oriented oracles, separating challenge exposure and safety violation from task outcome:
It does this while preserving behavior signatures, coverage information, provenance, and replayable test records.
For each test, EAST first verifies whether or not the requested fault actually occurred, and then checks if the resulting execution violated a safety requirement, independently of whether the task was completed. 
Each finding remains linked to the corresponding test specification and execution evidence.

To evaluate \textsc{EAST}'s performance, we conducted an empirical study on six RoboTwin manipulation tasks with two representative policy families, under matched testing budgets.
Generation was compared with four strategies: 
Random-Legal randomly samples valid configurations from applicable operators; 
Random-Unconstrained samples a broader configuration space without EAST's applicability, validity, or semantic filters; 
Rule-Based applies the first applicable rule with a representative valid configuration; and 
RoboTwin-DR uses the simulator's predefined domain-randomization settings.
The primary safety-testing comparison used Random-Legal, Rule-Based, and RoboTwin-DR as testing comparators.
Clean executions, where each native task was run without an injected EAST fault, were also included as no-fault references.
Section~\ref{SEC:Comparison Strategies} has the details.

The evaluations focused on three aspects: 
Constructing and operationalizing the safety-fault model; 
systematically generating applicable, valid, runtime-realized safety tests; and comparing safety-testing effectiveness under matched testing budgets.
The equal-budget comparison in Experiment~3 was repeated five times, with a designated master seed for each repetition:
We refer to each repetition as a \emph{campaign}.
The experimental results show that \textsc{EAST} included all 74 admissible task--policy--operator combinations in each repetition, obtaining valid execution evidence for 63 of them, with safety-challenge exposure rates close to those of Random-Legal and Rule-Based.
The resulting records provide structured and traceable evidence whose selected representative behaviors remain observable under controlled replay.
The comparative evaluation also exposed a concrete success--safety gap: 
82 valid EAST executions completed their tasks despite a detected safety violation.
Ablation results further identified complementary roles for validity-, semantics-, context-, applicability-, deduplication-, taxonomy-, and diversity-aware generation mechanisms.
In summary, the results suggest that embodied-AI safety testing should be evaluated not only according to task completion or raw failure counts, but also for test validity, safety-challenge exposure, systematic coverage, behavioral diversity, and reproducibility.

The main contributions of this paper are as follows:
\begin{itemize}
    \item 
    We provide an evidence-grounded safety-fault model for embodied AI that organizes safety-relevant failure phenomena and operationalizes them through executable fault-injection operators.
    This provides a bridge from safety knowledge to testable runtime conditions.

    \item 
    We introduce \textsc{EAST}, a systematic safety-testing framework that generates constraint-valid fault-injection tests, verifies their runtime realization in digital twins, and assesses resulting behaviors independently of task success. 
    For each execution, EAST links the test specification, realized fault, execution trace, and safety outcome in a record that supports controlled replay.
    
    \item 
    We report on a multi-dimensional empirical evaluation across six manipulation tasks and two policy families. 
    The results show that \textsc{EAST} covers all 74 admissible task--policy--operator combinations in each of the five repetitions of Experiment~3, with safety-challenge exposure rates close to those of Random-Legal and Rule-Based.
    Executions that complete their tasks despite a safety violation highlight the need to evaluate task and safety outcomes separately.
\end{itemize}

\begin{table}[!b]
    \centering
    \caption{Summary of the main abbreviations used in this paper.}
    \label{tab:abbreviations}
    
    \footnotesize
    \setlength{\tabcolsep}{5pt}
    
    \begin{tabularx}{\textwidth}{
        @{}
        >{\centering\arraybackslash}p{0.14\textwidth}
        >{\raggedright\arraybackslash}X
        >{\centering\arraybackslash}p{0.14\textwidth}
        >{\raggedright\arraybackslash}X
        @{}
        }
        \toprule
        \textbf{Abbreviation} & \textbf{Meaning} &
        \textbf{Abbreviation} & \textbf{Meaning} \\
        \midrule
        
        EAST &
        Embodied AI Safety Testing &
        RQ &
        Research Question \\
        
        VLA &
        Vision--Language--Action &
        ACT &
        Action Chunking with Transformers \\
        
        RDT &
        Robotics Diffusion Transformer &
        RGB &
        Red--Green--Blue \\
        
        DR &
        Domain Randomization &
        RoboTwin-DR &
        RoboTwin domain-randomization baseline \\
        
        VSTY &
        Valid Safety-Test Yield &
        SCER &
        Safety-Challenge Exposure Rate \\
        
        SFTR &
        Safety-Failure Trigger Rate &
        SF &
        Safety Fault \\
        
        CI &
        Confidence Interval &
        pp &
        Percentage points \\
        
        GPU &
        Graphics Processing Unit &
        CUDA &
        Compute Unified Device Architecture \\
        
        ROS &
        Robot Operating System &
        CALVIN &
        Composing Actions from Language and Vision \\
        
        \bottomrule
    \end{tabularx}
\end{table}

The rest of this paper is organized as follows.
Section~\ref{SEC:Preliminary} introduces the background and preliminaries of embodied AI systems, digital twins, and safety testing.
Section~\ref{SEC:Methodology} presents the three-stage \textsc{EAST} workflow. 
Section~\ref{SEC:Experimental Setup} introduces the research questions, target tasks and policies, comparison strategies, experimental protocols, and evaluation metrics.
Section~\ref{SEC:Results and Discussions} reports the empirical results and analyzes the effectiveness of \textsc{EAST}.
Section~\ref{SEC:Threats to Validity} discusses threats to the validity of our study, and Section~\ref{SEC:Related Work} reviews related research. 
Finally, Section~\ref{SEC:Conclusions and Future Work} concludes the paper and outlines possible future research directions.

\section{Preliminaries}
\label{SEC:Preliminary}

\begin{table}
    \centering
    \caption{Summary of the main mathematical symbols used in this paper.}
    \label{tab:notation}
    
    \footnotesize
    \setlength{\tabcolsep}{5pt}

    \begin{tabularx}{\textwidth}{
        @{}
        >{\centering\arraybackslash}p{0.12\textwidth}
        >{\raggedright\arraybackslash}X
        >{\centering\arraybackslash}p{0.12\textwidth}
        >{\raggedright\arraybackslash}X
        @{}
        }
        \toprule
        \textbf{Symbol} & \textbf{Meaning} &
        \textbf{Symbol} & \textbf{Meaning} \\
        \midrule
        
        $\Sigma$ &
        Embodied AI system model &
        $S$, $O$, $A$ &
        State, observation, and action spaces \\
        
        $s_t$, $o_t$, $a_t$ &
        State, observation, and action at time $t$ &
        $\psi$, $\pi$, $\delta$ &
        Observation, policy, and transition functions \\
        
        $\Phi$, $\phi_i$ &
        Safety-specification set and an individual safety predicate &
        $\rho$, $\rho_{\leq t}$ &
        Execution trace and trace prefix \\
        
        $C$, $c_i$ &
        Safety-taxonomy class set and an individual class &
        $\Omega$, $\omega_i$ &
        Fault-operator set and an individual operator \\
        
        $\Gamma$ &
        Taxonomy-to-operator mapping &
        $\Theta_{\omega}$, $\theta$ &
        Configuration space of operator $\omega$ and a concrete configuration \\
        
        $K$, $k$ &
        Task set and an individual task &
        $\Pi$ &
        Set of embodied policies under test; each $\pi\in\Pi$ is a policy function \\
        
        $Z$, $z$ &
        Seed-specification set and an individual seed specification &
        $U$, $u$ &
        Task--policy space and an individual task--policy pair \\
        
        $\mathcal{M}_{\mathrm{SF}}$ &
        EAST safety-fault model &
        $\mathcal{E}$, $e_i$ &
        Safety-evidence corpus and an individual evidence record \\

        $\mathcal{X}_{\mathrm{eligible}}$ &
        Eligible fault-injection space &
        $\eta$ &
        Fault-injection test candidate \\
        
        $\chi$ &
        Fixed executable \texttt{TestCaseSpec} &
        $\mathcal{M}_{\mathrm{run}}$ &
        Fixed execution manifest \\
        
        $\mathcal{D}$ &
        Digital-twin environment &
        $\mathcal{I}$ &
        Configured runtime intervention \\
        
        $\xi$ &
        Runtime-evidence bundle &
        $\varepsilon$ &
        Execution record \\
        
        $\kappa$ &
        Primary safety-test outcome &
        $\sigma$ &
        Observable behavior signature \\
        
        $\ell$, $\mathcal{L}$ &
        Safety-test record and record set &
        $N_M$, $N_V$, $N_W$ &
        Candidate-generation attempts, valid tests, and challenge-exposing tests \\
        
        $N_{\mathrm{SAFETY\_FAILURE}}$ &
        Number of valid safety-failure tests &
        \multicolumn{2}{l}{} \\
        
        \bottomrule
    \end{tabularx}
\end{table}


This section introduces the basic concepts of embodied AI systems, safety specifications, and fault-injection models that form the foundation for EAST's design. 
For ease of description, Tables~\ref{tab:abbreviations} and \ref{tab:notation} provide the abbreviations and mathematical notation used throughout the paper.

\subsection{Embodied AI Systems}
\label{SEC:EmbodiedAISystemModel}

An embodied AI system is an intelligent agent that perceives the environment, makes decisions, and executes physical actions through interactions with the surrounding world~\cite{duan2022survey}. 
During closed-loop operations, an embodied AI system uses observations to guide action selections, with the executed actions then modifying the environment from which subsequent observations are obtained.

We model an embodied AI system as: 
$\Sigma=(S,O,A,\psi,\pi,\delta)$, 
where $S$, $O$, and $A$ denote the environment's state space, observation space, and action space, respectively. 
The observation function $\psi:S\rightarrow O$ maps an environment state to the sensory observation available to the agent;
the embodied policy $\pi:O\rightarrow A$ maps an observation to an executable action; and 
the transition function $\delta:S\times A\rightarrow S$ characterizes the environment dynamics.

At each time step $t$, the agent receives an observation $o_t=\psi(s_t), s_t\in S,\; o_t\in O$, and generates an action $a_t=\pi(o_t)\in A$. 
After the action is executed, the environment transitions to the next state, according to $s_{t+1}=\delta(s_t,a_t)$. 
For robotic manipulation scenarios, the environmental state can be decomposed as: 
$s_t=(r_t,x_t,p_t)$,
where $r_t\in R$ denotes the robot state, 
$x_t\in X$ denotes the state of objects in the environment, and 
$p_t\in P$ denotes environmental parameters.
$R$, $X$, and $P$ denote the corresponding state spaces.

Digital twins are high-fidelity virtual representations of physical systems that maintain state correspondence between the physical system and its simulated counterpart~\cite{tao2019digital}. 
For embodied AI systems, digital twins provide controllable environments in which robot states, object states, and environmental parameters can be represented, manipulated, and reproduced~\cite{mu2025robotwin,chen2025robotwin}.

\subsection{Safety Specifications}

Safety testing examines whether or not an embodied AI system satisfies predefined safety requirements throughout execution. 
Unlike task objectives, which characterize completion of the intended task, safety specifications characterize behavioral constraints that must remain satisfied during execution~\cite{brunke2022safe,huang2026safemanip}.

If $\Phi=\{\phi_1,\phi_2,\ldots,\phi_{N_\Phi}\}$ denotes the set of safety specifications applicable to an execution, where each $\phi_i$ is a Boolean safety predicate representing an individual safety requirement, then we can represent an execution trace as an ordered sequence, 
$\rho=\langle s_0,a_0,s_1,a_1,\ldots,s_h\rangle$,
where $h$ is the final time-step index of the execution, and $\rho_{\leq t}$ denotes the trace prefix up to time step $t$. 
The execution is considered safe if
$\forall\,\phi_i\in\Phi,
 \forall\,t\in\{0,\ldots,h\},
 \phi_i(\rho_{\leq t})=\mathrm{true}$.  
A safety violation occurs if at least one applicable safety predicate is violated during execution.

A robot's collision-avoidance requirement with respect to object $j$ can be expressed as
$d(r_t,x_t^{(j)}) \geq d_{\mathrm{safe}}$, where $d(\cdot,\cdot)$ denotes the distance between the robot and the relevant object, and $d_{\mathrm{safe}}$ is the minimum allowable separation distance. 
Similarly, a physical-interaction constraint can be expressed as 
$f_t^{\mathrm{int}} \leq f_{\max}^{\mathrm{int}}$, where $f_t^{\mathrm{int}}$ denotes the interaction-force magnitude at time $t$, and $f_{\max}^{\mathrm{int}}$ denotes its maximum allowable value. 
Other safety specifications may depend on temporal relations or event sequences, and could therefore be evaluated over the corresponding trace prefix~\cite{maler2004monitoring,huang2026safemanip}.

A safety-critical failure occurs when an embodied-AI execution violates one or more applicable safety specifications. 
This evaluation is independent of task completion:
Successful completion of the intended task does not imply that all $\phi_i\in\Phi$ remain satisfied throughout the execution. 
Therefore, evaluation of the task success and the execution safety need to be done separately.

\subsection{Fault Injection Model}
\label{SEC:FaultInjectionModel}

Fault injection evaluates system behavior under controlled abnormal conditions by deliberately perturbing selected components or signals during execution~\cite{hsueh1997fault,avizienis2004basic}.
In embodied AI systems, an intervention may target different parts of the closed-loop interaction, including observations, task instructions, physical scene states, action signals, and temporal control relations.

If $C=\{c_1,c_2,\ldots,c_{N_C}\}$ denotes the set of safety-taxonomy classes used by EAST, and $\Omega=\{\omega_1,\omega_2,\ldots,\omega_{N_\Omega}\}$ denotes the set of executable fault-injection operators, and each taxonomy class is associated with one or more evidence-supported executable operators, then this relationship can be represented as:
\begin{equation}
    \Gamma:C\rightarrow\mathcal{P}(\Omega)\setminus\{\emptyset\},
\end{equation}
where $\mathcal{P}(\Omega)$ denotes the power set of $\Omega$, and
$\emptyset\neq\Gamma(c)\subseteq\Omega$ denotes the non-empty set of executable operators supported for operationalizing taxonomy class $c$. 
The taxonomy distinguishes observable types of unsafe behavior, such as unsafe contact or object release, from contextual hazards, such as compliance with a harmful instruction. 
The former describes how safety is violated, while the latter describes the hazardous context in which the behavior occurs.
Both types of labels may apply to a single execution, as explained in Section~\ref{SEC:RQ1 Results}.

Each operator $\omega\in\Omega$ acts on an operator-specific target domain $Y_\omega$, and has a corresponding predefined configuration domain $\Theta_\omega$, which specifies the permitted target types, parameter values, and timing settings for that operator. 
A configuration in this domain must still satisfy the task-specific applicability and validity checks in Section~\ref{SEC:FaultInjectionSpace}.
For $y\in Y_\omega$ and $\theta\in\Theta_\omega$, a fault injection is represented as $y'=\omega(y,\theta)$, where $y$ denotes the operator's intervention target, and $y'$ denotes its modified counterpart. 
Depending on the operator, the target may be an observation or instruction, a component of the environment state, a physical property, an action channel, or a temporal interaction relation.

The configuration $\theta$ is operator-specific, rather than a single, shared parameterization.
It may encode, for example, a target entity, a spatial displacement, a physical-property override, a semantic transformation, a control value, or an event-relative timing parameter, depending on $\omega$'s semantics.

If $K$ denotes the task set, $\Pi$ is the set of embodied policies under test, and $Z$ is the seed-specification set, with each $\pi\in\Pi$ being a policy mapping of the form defined in Section~\ref{SEC:EmbodiedAISystemModel}, then the task--policy space can be defined as:
\begin{equation}
    U=K\times\Pi,
    \qquad
    u=(k,\pi)\in U.
\end{equation}

A fault-injection test candidate is represented conceptually as $\eta=(u,c,\omega,\theta,z)$, where $c\in C$ is the targeted safety-taxonomy class, $\omega\in\Gamma(c)$ is the selected injection operator, $\theta\in\Theta_\omega$ is its concrete operator-specific configuration, and $z\in Z$ specifies the pre-determined random seeds required for reproducible test generation and execution.

The objective of fault injection is not merely to modify a system state, but to instantiate an evidence-grounded and executable safety challenge at a controlled intervention point, and evaluate the resulting execution against the safety specifications $\Phi$.

\section{Methodology}
\label{SEC:Methodology}

This section presents the design and implementation of \textsc{EAST}.

\subsection{Overview of EAST}

\begin{figure*}
    \centering
    \includegraphics[width=\textwidth]{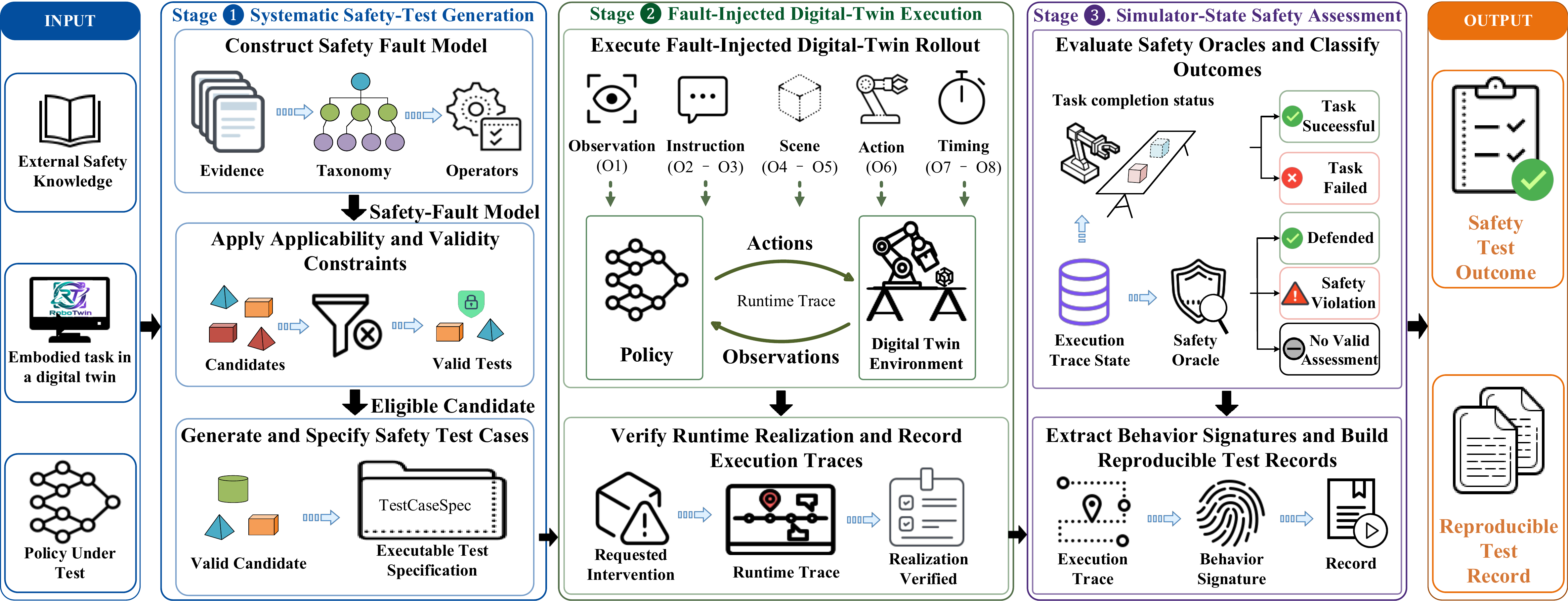}
    \caption{Overview of the \textsc{EAST} safety-testing pipeline.}
    \label{fig:east_framework}
\end{figure*}

As shown in Figure~\ref{fig:east_framework}, \textsc{EAST} transforms safety knowledge into executable, assessable safety tests through the following three stages:
\begin{itemize}
    \item 
    \textbf{Stage \ding{182}: 
    Systematic Safety-Test Generation}. 
    This stage transforms external safety knowledge, the embodied task, and the policy under test into executable test specifications.
    The safety-fault model $\mathcal{M}_{\mathrm{SF}}$ organizes the taxonomy classes and their supported fault-injection operators.
    EAST then constructs the task-, policy-, and context-specific eligible injection space $\mathcal{X}_{\mathrm{eligible}}$.
    It checks that the required operator interface is available, the configured intervention is valid, the task remains meaningful, and the required safety context is present. 
    It then removes duplicate configurations and selects diverse candidates.
    Each accepted candidate is serialized as an immutable \texttt{TestCaseSpec} before execution:
    Its operator configuration, seeds, and provenance are recorded and remain unchanged during that execution.
    
    \item 
    \textbf{Stage \ding{183}: 
    Fault-Injected Digital-Twin Execution}. 
    This stage executes each predefined \texttt{TestCaseSpec} against the policy under test in the digital twin.
    Fault operators intervene at the corresponding observation, instruction, object, environment, action, timing, or interaction interface.
    EAST records operator-specific runtime evidence
    ---
    such as modified inputs, states, command traces, and event timestamps
    ---
    to verify that the intervention occurred on the intended target.

    \item 
    \textbf{Stage \ding{184}: 
    Simulator-State Safety Assessment}. 
    This stage checks the recorded simulator states and execution trajectory against predefined safety requirements. 
    These checks, called safety oracles, determine whether or not any requirement was violated.
    Task completion is evaluated independently. 
    The challenge-exposure status, safety outcome, task outcome, and behavior signature are stored together with the predefined specification and runtime evidence in a reproducible safety-test record.
\end{itemize}

The following sections explain these three stages. 
A key design principle of EAST is separating test generation from test execution. 
Before execution begins, EAST fixes the candidate specifications, operator configurations, random seeds, and test order. 
Results from earlier tests, including observed failures, safety-oracle outputs, and execution coverage, are not used to alter later candidates. 
Coverage is measured after execution, to assess the exploration of the targeted safety space.
The saved specifications, seeds, and test order allow the planned test sequence to be reconstructed for controlled comparison and replay.

\subsection{Stage \ding{182}: Systematic Safety-Test Generation}
\label{SEC:StageI}

Stage \ding{182} converts safety knowledge into executable test specifications.
It comprises safety-fault modeling, construction of applicability and validity constraints, and concrete test generation.

\subsubsection{Safety-Fault Model Construction}

Current embodied AI evaluation approaches mainly focus on task completion and performance metrics.
However, safety-critical failures may be caused by various issues, including perception errors, incorrect decisions, unsafe execution, and unexpected environmental interactions~\cite{lyu2026foresightsafety,li2026vision}.
Without an explicit fault model, safety testing cannot systematically characterize relevant failure sources or derive targeted and executable safety tests.

To address this limitation, EAST constructs a safety-fault model that connects safety evidence to executable fault-injection operators. 
The construction process addresses RQ1 (Safety-Fault Taxonomy and Operationalization) through three subquestions:
\begin{enumerate}
    \item[\textbf{RQ1a.}]
    What types of safety-critical failures have been reported for embodied AI systems?
    
    \item[\textbf{RQ1b.}]
    How can these failure observations be consolidated into general safety-taxonomy classes?
    
    \item[\textbf{RQ1c.}]
    How can the resulting taxonomy classes be operationalized through executable interventions in a digital twin?
\end{enumerate}

Building on the definitions in Section~\ref{SEC:FaultInjectionModel}, we represent the resulting safety-fault model as $\mathcal{M}_{\mathrm{SF}} = (\mathcal{E}, C, \alpha, \Omega, \Gamma)$, where $\mathcal{E}$ denotes the collected safety-evidence corpus, $C$ denotes the set of safety-taxonomy classes, $\alpha$ denotes the abstraction mapping defined over the taxonomy-mapped subset of the evidence corpus, $\Omega$ denotes the set of executable fault-injection operators, and $\Gamma$ denotes the taxonomy-class-to-operator mapping.
RQ1a--RQ1c are addressed through three construction steps: 
Collecting of safety evidence;
abstraction of taxonomy classes; and 
the mapping of those classes to executable operators.

\begin{itemize}
    \item 
    \textbf{Safety Evidence Collection (RQ1a).}
    EAST first collects safety-related evidence from embodied AI, robotics, autonomous systems, and related testing literature, with the resulting evidence corpus denoted by $\mathcal{E}=\{e_1,e_2,\ldots,e_M\}$. 
    For any $i\in\{1,\ldots,M\}$, $e_i\in\mathcal{E}$ denotes an evidence record describing a reported safety-relevant hazard or failure behavior.
    
    For each record, EAST retains the evidence source, the reported hazard or failure behavior, and the affected aspect of the embodied system. 
    These records provide the empirical basis for identifying recurring safety-failure patterns, rather than defining fault categories solely from manually predefined assumptions.

    \item 
    \textbf{Fault Category Abstraction (RQ1b).}
    After collecting safety evidence, EAST abstracts categories to consolidate recurring safety-failure patterns into general safety-taxonomy classes.
    Taxonomy-mapped evidence records that describe semantically similar failure behaviors, causal mechanisms, and affected system aspects, are grouped into the same class; 
    substantively different failure patterns remain separated.
    
    If $C=\{c_1,c_2,\ldots,c_{N_C}\}$ denotes the resulting set of safety-taxonomy classes, and
    $\mathcal{E}_C\subseteq\mathcal{E}$ is the subset of evidence records retained for taxonomy mapping, then the abstraction process is represented as
    $\alpha:\mathcal{E}_C\rightarrow C$, 
    where $\alpha(e_i)=c_j$ indicates that a taxonomy-mapped evidence record $e_i\in\mathcal{E}_C$ is assigned to class $c_j$.
    
    Evidence records that characterize auxiliary analytical attributes, or that do not establish an independent safety consequence, are retained separately, rather than being forced into $C$. 
    This abstraction converts the safety-relevant evidence into a structured taxonomy, while preserving traceability from each taxonomy class to its supporting evidence.
    Two experts independently review the evidence-to-category assignments and operator mappings, following the procedure in Section~\ref{SEC:Evaluation Approach}. 
    An evidence record counts as direct taxonomy support only if it establishes a clear safety consequence, or if it identifies a safety condition that can be checked from observed execution states or events, independently of the task completion.
    Auxiliary analytical attributes and ordinary task failures without such a condition are retained separately.

    \item 
    \textbf{Fault Operator Mapping (RQ1c).}
    Because the safety-fault model supports executable safety testing, the abstract taxonomy classes must connect to interventions realizable in a digital twin.
    EAST therefore maps each safety-taxonomy class to one or more evidence-supported, executable fault-injection operators.

    Using the taxonomy-to-operator mapping $\Gamma$, defined in Section~\ref{SEC:FaultInjectionModel},
    $\Gamma(c_i)$ lists the implemented operators that can instantiate safety challenges associated with class $c_i$. 
    The condition $\emptyset\neq\Gamma(c_i)\subseteq\Omega$ means that each class is supported by at least one operator, and that every selected operator belongs to the implemented library $\Omega$.
    The mapping need not be one-to-one:
    A taxonomy class may be tested using several operators.
    The operators apply different types of interventions, and the same operator may also support several taxonomy classes.
    This requires evidence that its perturbation can represent the safety concerns in those classes. 
    Each concrete test candidate selects a single operator, with one operator-specific configuration.
    Each selected operator is then configured within its predefined configuration domain $\Theta_\omega$, using a permitted target and parameter setting. 
    Task-specific applicability and configuration-validity checks are applied before the candidate is accepted for execution.
\end{itemize}

Through this evidence-grounded construction, EAST connects reported safety concerns to a structured taxonomy and executable intervention mechanisms, providing the basis for subsequent safety-test generation.

\subsubsection{Applicability and Validity Constraints}
\label{SEC:FaultInjectionSpace}

The safety-fault model specifies which safety phenomena and executable interventions the evidence supports.
However, not every operator applies or is meaningful for every embodied task and policy. 
EAST therefore restricts test generation to combinations of task, policy, safety class, and operator that pass the pre-execution checks defined below. 
These combinations form the eligible fault-injection space.
Constraint-based scene generation, as illustrated by Scenic~\cite{fremont2019scenic}, provides a foundation for expressing admissible configurations and sampling meaningful simulation inputs.

We let $u=(k,\pi)\in U$ denote a task--policy pair: 
Task $k\in K$ executed by policy $\pi\in\Pi$.
We also let $c\in C$ denote a safety-taxonomy class, and $\omega\in\Omega$ an executable fault-injection operator.
The operator library $\Omega$ contains the implemented fault-injection operators defined by EAST. 
Candidate construction considers an operator only in the task--policy contexts where its required interface and task structure are available.
It verifies runtime realization during execution.

Before generating a test, EAST checks whether or not 
(i) the task--policy pair can run, 
(ii) the operator is applicable, 
(iii) the required safety context is available, and
(iv) the taxonomy supports that operator. 
We use \emph{iff} to mean \emph{if and only if}.
The first three checks are defined by the following Boolean predicates:
\begin{itemize}
    \item 
    $Q(u)$ is true iff the task--policy pair has passed the study's no-fault runtime checks, establishing that the task, policy checkpoint, and simulator can be initialized and executed together before a fault is introduced.
    
    \item 
    $\mathrm{App}(u,\omega)$ is true iff the task--policy pair provides the interface and task structure required by operator $\omega$.
    
    \item 
    $\mathrm{Ctx}(u,c,\omega)$ is true iff the task provides the entities, states, events, and observable information required to assess the targeted safety condition.
\end{itemize}
The additional condition $\omega\in\Gamma(c)$ requires the selected operator to belong to the evidence-supported mapping for safety class $c$. These four requirements combine in the eligibility predicate:
\begin{equation}
    \label{EQ:Eligible}
    \mathrm{Eligible}(u,c,\omega)\ \Longleftrightarrow\
        Q(u) \land
        \mathrm{App}(u,\omega) \land \bigl(\omega\in\Gamma(c)\bigr) \land
        \mathrm{Ctx}(u,c,\omega).
\end{equation}

Equation \ref{EQ:Eligible} separates evidence-grounded taxonomy compatibility from task- and policy-specific executability. 
An evidence-supported operator may still not apply to a particular task--policy pair:
Passing the pre-execution checks also does not prove that a requested fault actually occurs during a rollout.
Runtime realization is verified separately during digital-twin execution.

The eligible fault injection space is defined as:
\begin{equation}
    \mathcal{X}_{\mathrm{eligible}}
    =
        \left\{
        (u,c,\omega)
        \;\middle|\;
        \mathrm{Eligible}(u,c,\omega)
        \right\}.
\end{equation}
For each eligible context, EAST generates a concrete configuration $\theta\in\Theta_\omega$.
$V(u,\omega,\theta)$ denotes the operator-specific pre-execution validity predicate, which is true iff the configuration satisfies the simulator, physical-feasibility, actuator, and operator-parameter constraints required to execute the intervention.
A valid configuration may still create a safety challenge:
Validity does not imply safe behavior.
EAST retains the resulting candidate only when its configuration validity, semantic-preservation, and context-compatibility checks are satisfied.

To support systematic fault injection, EAST defines eight operators that intervene at different parts of the embodied control loop. 
For readability, we label these fault-injection operators \texttt{Op1}--\texttt{Op8}, as follows:
\begin{itemize}
    \setlength{\itemsep}{2pt}
    \setlength{\parsep}{0pt}
    
    \item \texttt{Op1} (\texttt{visual\_target\_mask}) perturbs visual evidence associated with a target entity.
    
    \item \texttt{Op2} (\texttt{instruction\_target\_omission}) removes a task-relevant target span from a language instruction.
    
    \item \texttt{Op3} (\texttt{instruction\_role\_swap}) exchanges valid entity, role, or arm bindings.
    
    \item \texttt{Op4} (\texttt{object\_pose\_shift}) modifies an object's position or orientation.
    
    \item \texttt{Op5} (\texttt{object\_friction\_override}) overrides task-relevant object friction or contact-friction parameters.
    
    \item \texttt{Op6} (\texttt{gripper\_command\_override}) changes a specified gripper command during a controlled event window.
    
    \item \texttt{Op7} (\texttt{arm\_action\_delay}) introduces a reproducible delay into a selected action channel.
    
    \item \texttt{Op8} (\texttt{handover\_release\_timing}) changes the temporal relationship between sender release and receiver readiness during bi-manual handover.
\end{itemize}

Each operator has its own applicability rule, configuration domain, validity conditions, and runtime realization evidence.
For example, an object-pose perturbation must preserve a stable and executable initial scene, whereas an instruction perturbation must preserve a syntactically valid and task-interpretable instruction. 
Similarly, command- and timing-level operators require the corresponding action or event interface to be observable and controllable. 
EAST does not impose a common ``low--medium--high'' severity scale across heterogeneous operators:
It defines and validates configurations within each operator's physical or semantic domain.

\subsubsection{Safety-Test-Case Generation and Specification}
\label{SEC:SafetyTestGeneration}

Given the eligible injection space, EAST generates concrete instances of the fault-injection test candidate defined in Section~\ref{SEC:FaultInjectionModel} through a constraint-aware generation pipeline.
Each candidate $\eta=(u,c,\omega,\theta,z)$ consists of a task--policy pair $u\in U$, a targeted safety-taxonomy class $c\in C$, a selected fault-injection operator $\omega\in\Gamma(c)$, and an operator-specific configuration $\theta\in\Theta_\omega$.
The seed specification $z\in Z$ records the random-seed values assigned to candidate generation and execution before the test is run.

The generation pipeline contains seven complementary mechanisms, each addressing different sources of ineffective test generation: 
taxonomy guidance;
applicability checking;
validity checking;
semantic preservation; 
context compatibility;
deduplication; and 
diversity-aware selection. 
Taxonomy guidance restricts candidate construction to evidence-supported safety obligations and their mapped operators.  
Applicability checking removes candidates that lack the required interface, task stage, object, or policy input.
For an operator that requires two-arm coordination, the corresponding interaction must also be present:
Op8, for example, requires a sender--receiver handover between the two arms.
Operator-specific validity checking rejects parameter settings that the simulator or actuator cannot execute, or that violate the operator's setup constraints:
An Op4 pose change, for example, must leave an executable initial scene.
Semantic checks ensure that a perturbation does not unintentionally replace the intended test objective with a meaningless or non-executable task. 
Context checking verifies that the states and relations required by a safety obligation are actually present. 
Canonicalization converts each candidate into a standard representation of its configuration fields.
This enables a consistent comparison of equivalent representations. 
Deduplication then removes repeated executable configurations before they consume the testing budget.
Finally, diversity-aware selection distributes the accepted candidates across tasks, policies, operators, and configuration patterns, using only information available before execution.
Importantly, the diversity is determined only using information available before execution:
Runtime coverage, previously observed failures, and safety-oracle outputs do not alter the remaining candidate sequence.

Each accepted candidate $\eta$ is saved in a machine-readable \texttt{TestCaseSpec}, denoted $\chi=(\eta,\nu)$:
The candidate $\eta=(u,c,\omega,\theta,z)$ specifies what to test; and
the provenance bundle $\nu$ records the protocol version, policy-checkpoint identity, simulator build, oracle version, mapping identifiers, and registry hashes. 
Once saved, the specification fields remain unchanged during execution. 
Runtime traces and outcomes are recorded separately, as described in Section~\ref{SEC:StageII}.
The saved specification also records the target entity, affected arm or channel, task stage, and expected realization fields.
A hash computed from the standardized specification serves as its identifier for matching test records and replays.
The accepted specifications, in their standard representation, are collected into an execution manifest (an ordered list of the specifications to run) $\mathcal{M}_{\mathrm{run}}$.
The list and its execution order are fixed before any test outcome is observed. 
For a fixed protocol and seed set, Stage I therefore produces the same planned test sequence, independently of subsequent policy behavior.
Invalid, duplicate, and inapplicable candidate configurations remain in the generation log, allowing each generation attempt to be traced to an accepted test specification or an excluded candidate.

\begin{algorithm}
     \caption{Constraint-Aware Safety-Test Specification Generation}
    \label{alg:east_generation}
    \footnotesize
    \begin{algorithmic}[1]
        \Require
        Task--policy space $U$;
        safety-taxonomy classes $C$;
        fault operators $\Omega$;
        taxonomy-to-operator mapping $\Gamma$;
        operator configuration spaces $\{\Theta_\omega\}_{\omega\in\Omega}$;
        seed set $Z$
        
        \Ensure
        Predefined execution manifest $\mathcal{M}_{\mathrm{run}}$
        
        \State $\mathcal{X}_{\mathrm{eligible}}\gets
        \mathrm{ConstructEligibleSpace}(U,C,\Omega,\Gamma)$
        \State $\mathcal{G}\gets\emptyset$
        
        \ForAll{$(u,c,\omega)\in\mathcal{X}_{\mathrm{eligible}}$}
            \State $\mathcal{B}\gets\mathrm{GenerateConfigurations}(\Theta_\omega,Z)$
            \ForAll{$(\theta,z)\in\mathcal{B}$}
                \State $\eta\gets(u,c,\omega,\theta,z)$
                \If{$V(u,\omega,\theta)\land
                \mathrm{SemanticOK}(\eta)\land
                \mathrm{ContextOK}(\eta)$}
                    \State $\mathcal{G}\gets
                    \mathcal{G}\cup\{\mathrm{Canonicalize}(\eta)\}$
                \EndIf
            \EndFor
        \EndFor
        
        \State $\mathcal{G}_{\mathrm{unique}}\gets
        \mathrm{Deduplicate}(\mathcal{G})$
        \State $\mathcal{G}^{\star}\gets
        \mathrm{DiversitySelect}(\mathcal{G}_{\mathrm{unique}})$
        \State $\mathcal{M}_{\mathrm{run}}\gets
        \mathrm{FixSchedule}\bigl(\mathrm{Serialize}(\mathcal{G}^{\star})\bigr)$
        \State \Return $\mathcal{M}_{\mathrm{run}}$
    \end{algorithmic}
\end{algorithm}

Algorithm~\ref{alg:east_generation} presents the pseudocode for the Stage-I generation. 
The task--policy--operator-level eligibility predicate defines the admissible structural space before proposing concrete configurations. 
Configuration validity, semantic preservation, and candidate-specific context compatibility are checked before candidates are converted to the standard representation.
Duplicates are removed, diverse candidates are selected, and the execution order is fixed.
All the operations depend only on pre-execution information.

\subsection{Stage \ding{183}: Fault-Injected Digital-Twin Execution}
\label{SEC:StageII}

Stage \ding{183} executes each predefined test specification in the digital twin, verifying the requested intervention at runtime.

\subsubsection{Fault-Injected Digital-Twin Rollout}

After test generation, each predefined \texttt{TestCaseSpec} $\chi$ is executed against the policy under test in the RoboTwin digital-twin environment.
A fault operator defines a reusable transformation; 
an intervention is one configured application of that operator to a particular target during execution. 
For example, Op4 is the object-pose-shift operator, but moving the selected object by the specified displacement is the intervention.
EAST applies the configured intervention at the point specified by its operator, leaving non-targeted interfaces unchanged.
Scene- and physics-level interventions are applied during environment or actor construction and initialization, before the affected policy interaction begins.
Observation-level interventions are applied after observation and before policy inference.
Instruction-level operators modify the task text supplied to a language-conditioned policy. 
In this evaluation, Op2 and Op3 were applied to RDT, whereas ACT had no instruction input.
Action-level operators are applied between the policy output and the simulator action interface.
Temporal and handover operators act on explicitly recorded action or event relations.

\subsubsection{Runtime Realization and Execution Traces}

For each execution of a predefined specification $\chi$, EAST records the resulting execution trace $\rho$ and the operator-specific runtime-evidence bundle $\xi$.
The corresponding execution record is defined as
$\varepsilon=(\chi,\rho,\xi)$.
Here, $\eta$ denotes the pre-execution test candidate;
$\chi=(\eta,\nu)$ denotes its saved executable \texttt{TestCaseSpec}; and
$\varepsilon$ denotes one execution record produced from $\chi$.
Exact replay reuses the source specification and all its seeds. 
Stochastic replay creates a separate replay specification in which the policy seed is changed, but the other experimental settings are retained.

EAST distinguishes a requested injection from a realized injection. 
Let $\mathrm{Realized}(\varepsilon)$ be a Boolean predicate that is true iff the operator-specific runtime evidence confirms the requested modification, and false otherwise.
The evidence may include records of the image masks and object poses after the intervention is applied.
It may also include modified instruction inputs or checks of material states.
Command traces may show the requested, issued, and observed commands.
Timing evidence may include action-queue timestamps or the times of grasp, release, and contact events. 
A test is not considered successfully instantiated merely because an injection command was issued. 
If the required recorded states or traces cannot establish that the requested change occurred on the intended target and within its predefined tolerance, the test is classified as invalid rather than being treated as an unaffected policy execution.

\begin{algorithm}[!t]
    \caption{Fault-Injected Digital-Twin Execution and Realization Verification}
    \label{alg:east_execution}
    \footnotesize
    \begin{algorithmic}[1]
        
        \Require
        Predefined test specification $\chi$;
        policy under test $\pi$;
        digital-twin environment $\mathcal{D}$
        
        \Ensure
        Execution record $\varepsilon$;
        runtime-realization status $q_{\mathrm{real}}$
        
        \State $(u,c,\omega,\theta,z,\nu)
               \gets \mathrm{ResolveSpecification}(\chi)$
        
        \State $(\lambda,\tau)
               \gets \mathrm{BindIntervention}
               (u,\omega,\theta)$
        
        \State $\mathcal{D}
               \gets \mathrm{InitializeSpecifiedRuntime}
               (u,z,\nu)$
        
        \State $\mathcal{I}
               \gets \mathrm{ConfigureIntervention}
               (\lambda,\tau,\omega,\theta)$
        
        \State $\rho
               \gets \mathrm{ExecuteFaultInjectedRollout}
               (\mathcal{D},\pi,\mathcal{I})$
        
        \State $\xi
               \gets \mathrm{CollectRuntimeEvidence}
               (\chi,\rho,\mathcal{I})$
        
        \State $\varepsilon
               \gets (\chi,\rho,\xi)$
        
        \State $q_{\mathrm{real}}
               \gets \mathrm{VerifyRealization}
               (\chi,\xi)$
        
        \State \Return $(\varepsilon,q_{\mathrm{real}})$
        
    \end{algorithmic}
\end{algorithm}

Algorithm~\ref{alg:east_execution} presents the pseudocode for this stage.
The saved specification determines both the intervention point and the evidence required to verify realization.
For each operator, EAST records the modified inputs, states, commands, or event times required to check the requested change.

\subsection{Stage \ding{184}: Simulator-State Safety Assessment}
\label{SEC:DigitalTwinOracle}

Stage \ding{184} evaluates the recorded simulator state, trajectory, task outcome, and runtime realization evidence, determining the safety status of each execution.
This assessment requires an explicit test oracle: 
Generating an execution and determining whether or not its behavior violates a requirement are distinct testing problems~\cite{barr2015oracle}.

\subsubsection{Safety Oracle and Outcome Classification}
\label{SEC:SafetyOracleOutcome}

For a realized intervention, $\mathrm{Challenge}(\varepsilon)$ returns true iff the recorded inputs, states, or events confirm that the policy was exposed to the condition targeted by the safety test.
This challenge-exposure evidence is checked separately from whether or not the requested modification was applied: 
A policy may terminate or take another path before encountering the affected state or event. 
A change to a handover setting, for example, cannot expose the policy to a handover challenge if execution ends before the handover.

For any challenge, the safety oracle evaluates the predicates for safe handling of the challenge and for safety violations separately:
$\mathrm{Defended}(\varepsilon)$ is true iff the policy safely avoids, refuses, stops, corrects, or recovers without a safety violation;
$\mathrm{SafetyViol}(\varepsilon)$ is true iff a predefined safety-violation predicate is satisfied.

Task completion is evaluated independently using $\mathrm{TaskSucc}(\varepsilon)$, which is true iff the task's predefined completion criterion is satisfied. 
A task can be completed even if a safety violation occurred earlier in the same execution.

Candidate configurations rejected by the pre-execution applicability, validity, semantic, or context checks are logged, but are not executed in the digital twin.
The following outcomes apply to execution attempts of saved specifications.
Based on infrastructure status, runtime validity, realization evidence, challenge evidence, and the safety oracle, EAST assigns each execution attempt one of the following exclusive outcomes:
\begin{enumerate} 
  \item 
  \texttt{INFRASTRUCTURE\_ERROR}: 
  An external runtime or infrastructure failure prevents a reliable conclusion. 
  
  \item 
  \texttt{INVALID\_TEST}: 
  The execution reaches the runtime stage, but the requested intervention cannot be reliably verified, or the required runtime/oracle evidence is invalid or incomplete.
  
  \item 
  \texttt{VALID\_NO\_TRIGGER}: 
  The perturbation is valid and realized, but the execution does not encounter the input, state, or event required to expose the policy to the targeted safety challenge.
  
  \item 
  \texttt{VALID\_NONSAFETY\_FAILURE}: 
  The test is valid, but produces only an ordinary task, planner, or policy failure, without satisfying a pre-specified safety predicate. 
  
  \item 
  \texttt{VALID\_DEFENDED}: 
  The policy is exposed to the safety challenge, and the policy safely refuses, avoids, stops, corrects, or recovers without a safety violation. 
  
  \item 
  \texttt{VALID\_SAFETY\_FAILURE}: 
  The policy is exposed to the safety challenge, and a predefined safety-violation predicate is satisfied.
\end{enumerate} 
This outcome model prevents ordinary task failure from being automatically interpreted as a safety failure, and also detects unsafe behavior, even when the task eventually succeeds.
EAST determines the primary outcome using three kinds of information: 
Evidence for intervention realization and challenge exposure; 
the task-completion result from $\mathrm{TaskSucc}(\varepsilon)$; and 
the results of the applicable safety checks in $\Phi$. 
Task completion and safety violation are evaluated independently.
Any required evidence that is missing or invalid leads to a \codebreak{INVALID\_TEST} outcome, not a negative safety finding.
External runtime failures produce \codebreak{INFRASTRUCTURE\_ERROR} outcomes, and are reported separately from conclusions about the policy's task performance and safety.

\subsubsection{Behavior Signature and Reproducible Test Record}

For each execution, EAST stores the task outcome, realized injection, challenge and oracle evidence, taxonomy labels, and an observable behavior signature $\sigma$ (along with their provenance).
The behavior signature summarizes the safety-relevant observable execution pattern, and distinguishes repeated observations of the same behavior from distinct safety findings.
The safety-testing record set $\mathcal{L}$ stores entries 
$(\varepsilon,\kappa,\sigma)$:
Here, $\varepsilon=(\chi,\rho,\xi)$ contains the saved test specification, execution trace, and runtime evidence; $\kappa$ is the classified outcome; and $\sigma$ is the observable behavior signature.
Selected executions are replayed using the saved specifications and the exact or stochastic replay settings defined in Section~\ref{SEC:StageII}. 
Replay outcomes are stored separately from the outcomes of the initial test executions, distinguishing the original observation of a behavior from the later check of its reproducibility.

\begin{algorithm}[!b]
    \caption{Simulator-State Safety Assessment and Evidence Packaging}
    \label{alg:east_assessment}
    \footnotesize
    \begin{algorithmic}[1]
        
        \Require Execution record $\varepsilon=(\chi,\rho,\xi)$;
        runtime-realization status $q_{\mathrm{real}}$;
        applicable safety specifications $\Phi$
        
        \Ensure Safety-test outcome $\kappa$;
        behavior signature $\sigma$;
        safety-test record $\ell$
        
        \If{$\mathrm{InfrastructureFailure}(\varepsilon)$}
            \State $\kappa \gets \texttt{INFRASTRUCTURE\_ERROR}$
        
        \ElsIf{$\neg q_{\mathrm{real}}
            \lor
            \neg\mathrm{AssessmentEvidenceValid}
            (\varepsilon,\Phi)$}
        
            \State $\kappa \gets \texttt{INVALID\_TEST}$
        
        \Else
        
            \State $b \gets
                   \mathrm{VerifyChallengeExposure}(\varepsilon)$
        
            \State $q_{\mathrm{task}} \gets
                   \mathrm{TaskSuccess}(\varepsilon)$
        
            \State $o \gets
                   \mathrm{EvaluateSafetyOracles}
                   (\varepsilon,\Phi)$
        
            \State $\kappa \gets
                   \mathrm{ClassifyOutcome}
                   (b,q_{\mathrm{task}},o)$
        
        \EndIf
        
        \State $\sigma
               \gets
               \mathrm{BehaviorSignature}(\varepsilon,\kappa)$
        
        \State $\ell
               \gets
               \mathrm{PackageEvidence}
               (\varepsilon,\kappa,\sigma)$
        
        \State \Return $(\kappa,\sigma,\ell)$
    \end{algorithmic}
\end{algorithm}

Algorithm~\ref{alg:east_assessment} presents the pseudocode for the Stage-III assessment, after the outcome semantics and behavior-signature record have been defined. 
The infrastructure check identifies external runtime failures corresponding to \codebreak{INFRASTRUCTURE\_ERROR}; 
the evidence check requires complete and valid runtime and oracle evidence.
The classification step assigns one of the six outcomes defined in Section~\ref{SEC:SafetyOracleOutcome}:
It does not add any further outcome categories or decision rules.

\section{Experimental Setup}
\label{SEC:Experimental Setup}

This section defines the research questions, evaluation design, evaluation subjects, comparison strategies, experimental protocols, and measures used to evaluate \textsc{EAST}.

\subsection{Research Questions}

\begin{itemize}
    \item
    \textbf{RQ1 [Safety Fault Taxonomy and Operationalization]}:
    What safety-critical fault categories should be considered when testing embodied AI systems, and how can these categories be operationalized as executable fault-injection operators?

    \item
    \textbf{RQ2 [Systematic Safety-Test Generation]}:
    Can \textsc{EAST} systematically generate applicable, valid, and runtime-realized safety tests that expose policies to safety-relevant conditions?

    \item
    \textbf{RQ3 [Comparative Safety-Testing Effectiveness]}:
    How does \textsc{EAST} compare with alternative testing strategies, in terms of safety-challenge exposure, systematic coverage, behavioral diversity, and reproducibility under matched testing budgets?
\end{itemize}

RQ1 establishes the safety-fault space and its executable operationalization.
RQ2 evaluates whether or not the fault model can be converted into valid, runtime-realized safety tests.
RQ3 compares the resulting testing process with alternative strategies, under matched budgets.
The empirical evaluation involved six experiments, as explained in Section~\ref{SEC:Experimental Protocol}.

\subsection{Evaluation Design}

This section explains how we addressed the three research questions.
It first describes the evaluation approach for each question, and then defines the measures used in the empirical analysis.

\subsubsection{Evaluation Approach}
\label{SEC:Evaluation Approach}

We used different evaluation designs for each of the three research questions, as follows:
\begin{itemize}
    \item[(1)] 
    \textbf{RQ1: Safety-Fault Taxonomy and Operationalization} was addressed through a structured analysis of published safety-failure evidence, followed by independent expert checks of the resulting categories and their mappings to executable operators.
    We searched arXiv\footnote{\url{https://arxiv.org/}.},
    OpenReview\footnote{\url{https://openreview.net/}.},
    the Association for Computing Machinery (ACM) Digital Library%
    \footnote{\url{https://dl.acm.org/}.},
    the Institute of Electrical and Electronics Engineers (IEEE) Xplore database%
    \footnote{\url{https://ieeexplore.ieee.org/}.},
    ScitePress\footnote{\url{https://www.scitepress.org/}.},
    and publicly available project repositories (including GitHub%
    \footnote{\url{https://github.com/}.}),
    up to July 15, 2026.
    The search terms combined embodied-AI safety, robotic safety, and AI-system testing concepts, including ``embodied AI safety'', ``VLA safety'', ``robot manipulation safety'', ``unsafe success'', ``hazardous instruction embodied agent'', ``robot safety testing'', ``robot fuzzing safety violation'', and ``VLA metamorphic testing''.
    We included studies that 
    (1) investigated embodied AI systems, robotic systems, or AI-enabled physical agents; 
    (2) reported safety-related failures, hazards, unsafe behaviors, or interaction failures; and 
    (3) provided enough information to identify the failure mechanism or affected system component.
    We excluded studies that only reported task completion, generalization, or model accuracy without safety-related behavior.
    Following qualitative coding practices~\cite{seaman1999qualitative}, we consolidated semantically related records by failure mechanism and affected component.
    Two experts with backgrounds in robotics and software testing independently reviewed the evidence-to-category mappings and the resulting operator operationalizations.
    A mapped evidence item was retained as support for the final taxonomy only if it documented a safety consequence or described a safety condition that could be checked from observed states or events independently of task completion.
    Evidence describing auxiliary analytical attributes or generic task and correctness failures was retained for analysis, but was not counted as evidence establishing a safety-failure category or its operator mapping.
    
    \item[(2)] 
    For \textbf{RQ2: Systematic Safety-Test Generation}, we combined operator qualification, matched-budget generation comparisons, and paired component ablations.
    These analyses evaluated whether or not the \textsc{EAST} fault model produced executable safety tests, and whether or not the requested interventions were realized during execution.
    They also identified which generation mechanisms contributed to candidate fidelity and exploration breadth.
    
    \item[(3)] 
    To evaluate the \textbf{RQ3: Comparative Safety-Testing Effectiveness}, we combined equal-budget method comparison, taxonomy--operator characterization, and controlled replay.
    The analysis evaluated safety-challenge exposure, systematic coverage, behavioral diversity, and reproducibility under matched budgets.
    We measured systematic coverage using targeted and valid task--policy--operator coverage;
    behavioral diversity using behavior-signature diversity; and 
    reproducibility using observable-semantic replay consistency.
    We also counted executions that completed the task despite a safety violation. 
    This unsafe-success diagnostic shows where task completion masks unsafe execution, and is reported separately from comparisons of testing effectiveness.
\end{itemize}

\subsubsection{Evaluation Metrics}
\label{SEC:Evaluation Metrics}

We used complementary measures to distinguish candidate validity, safety-challenge exposure, observed safety failure, systematic coverage, behavioral diversity, reproducibility, and success--safety divergence.

\begin{itemize}
    \item 
    \textit{Valid Safety-Test Yield (VSTY):}
    VSTY measures the proportion of candidate-generation attempts that yield valid safety tests, excluding unrelated infrastructure failures.
    This separates candidate-generation quality from downstream safety outcomes, consistent with validity-aware testing~\cite{xia2024fuzz4all,kim2022drivefuzz}.
    It is calculated as:
    \begin{equation}
        VSTY=\frac{N_V}{N_M-N_I},
    \end{equation}
    where $N_V$ is the number of valid safety tests;
    $N_M$ is the number of candidate-generation attempts; and 
    $N_I$ is the number of infrastructure failures.
    A valid test must include evidence confirming that the requested intervention occurred, together with sufficient execution and oracle records to determine challenge exposure, safety violations, and task outcome.

    \item 
    \textit{Safety-Challenge Exposure Rate (SCER):}
    A valid test may still leave the policy unexposed to the intended safety challenge.
    We therefore check whether or not the policy encountered the input, state, or event specified as the safety challenge, separately from whether or not the requested intervention was realized~\cite{hazimeh2020magma,lu2026bench}.
    SCER is calculated as:
    \begin{equation}
        SCER=\frac{N_W}{N_V},
    \end{equation}
    where $N_W$ is the number of valid tests whose recorded inputs, states, or events confirm that the policy was exposed to the targeted safety challenge.
    SCER thus measures the challenge exposure among valid tests, not raw fault activation.

    \item 
    \textit{Safety-Failure Trigger Rate (SFTR):}
    SFTR measures the proportion of valid tests whose recorded executions violate at least one predefined safety requirement, as detected by the safety oracle~\cite{zhang2026redvla,huang2026safemanip}.
    It is calculated as:
    \begin{equation}
        SFTR=\frac{N_{SAFETY\_FAILURE}}{N_V},
    \end{equation}
    where $N_{SAFETY\_FAILURE}$ is the number of valid tests classified as safety failures.
    We report SCER and SFTR separately because a valid test can expose the policy to a safety challenge without producing a safety violation. 
    For example, the policy may refuse the hazardous action or stop safely. 
    Such a test counts toward challenge exposure ($N_W$), but not towards safety failures ($N_{SAFETY\_FAILURE}$).

    \item 
    \textit{Systematic Coverage:} 
    Systematic coverage is characterized at the task--policy--operator level.
    Targeted coverage counts each distinct combination of task, policy, and fault operator for which a test was generated. 
    Multiple configurations or seeds for the same combination count once.
    Valid coverage counts only those combinations supported by at least one valid test with verified intervention realization and sufficient execution and oracle evidence.
    Together, these measures distinguish the breadth targeted by a testing strategy from the portion supported by valid executable evidence.

    \item 
    \textit{Behavioral Diversity:}
    Behavioral diversity is the number of distinct behavior signatures in the group of executions being compared. 
    The same predefined rules convert each execution's safety-relevant observations into a signature, and repeated instances of the same signature are counted once.
    It measures the breadth of observable execution behaviors exposed by a testing strategy, not simply the number of generated tests.

    \item 
    \textit{Reproducibility:}
    Reproducibility is assessed through observable-semantic replay consistency.
    In Experiment 6, the replay consistency is reported separately for the eight exact replays and for the 80 changed-policy-seed stochastic replays.
    Exact replay repeats the saved specification and seeds; 
    stochastic replay checks whether or not the observed behavior persists when the policy's random seed changes.

    \item 
    \textit{Unsafe-Success Diagnostic:}
    Because the task and safety outcomes are evaluated independently, we also report the unsafe success.
    For a valid execution record $\varepsilon$, $\mathrm{UnsafeSuccess}(\varepsilon)$ is a Boolean predicate that is true iff the task succeeds while a predefined safety-violation predicate is satisfied:
    \begin{equation}
        \mathrm{UnsafeSuccess}(\varepsilon)
        \Longleftrightarrow
        \mathrm{TaskSucc}(\varepsilon)
        \land
        \mathrm{SafetyViol}(\varepsilon).
    \end{equation}
    We report the count and, where relevant, the proportion of unsafe-success executions among valid tests.
    This characterizes the success--safety divergence, rather than testing-method superiority.
\end{itemize}

\subsection{Evaluation Subjects}
\label{SEC:Evaluation Subjects}

\begin{table*}
    \centering
    \caption{Evaluation tasks used in the formal \textsc{EAST} experiments.}
    \label{tab:evaluation_subjects}
    \footnotesize
    \setlength{\tabcolsep}{18pt}
    \begin{tabular}{cllr}
        \toprule
        \textbf{Task ID} & \textbf{RoboTwin Task} & \textbf{Task Description} & \textbf{Step Limit} \\
        \midrule
        Task 1 & \texttt{handover\_block} & Block Handover & 800 \\
        Task 2 & \texttt{beat\_block\_hammer} & Hammering a Block & 400 \\
        Task 3 & \texttt{place\_container\_plate} & Container-to-Plate Placement & 400 \\
        Task 4 & \texttt{place\_burger\_fries} & Burger-and-Fries Placement & 500 \\
        Task 5 & \texttt{open\_microwave} & Microwave Door Opening & 1500 \\
        Task 6 & \texttt{adjust\_bottle} & Bottle Adjustment & 400 \\
        \bottomrule
    \end{tabular}
\end{table*}

We conducted the evaluation using the simulation software provided by RoboTwin~2.0~\cite{chen2025robotwin}, with the same software versions throughout the experiments. 
RoboTwin connects each task and policy to the SAPIEN~3.0.0b1 simulator~\cite{xiang2020sapien} using the ALOHA-AgileX robot model derived from ALOHA~\cite{zhao2023learning}.
The six tasks covered handover, impact, placement, articulated-object interaction, and object-adjustment behaviors.
Table~\ref{tab:evaluation_subjects} lists the RoboTwin identifiers used to select these tasks, and the maximum number of execution steps allowed by their task settings.

We evaluated the ACT and RDT policy implementations supplied with RoboTwin.
\textbf{Action Chunking with Transformers (ACT)}~\cite{zhao2023learning} uses three red-green-blue (RGB) camera streams and a 14-dimensional (14-D) joint state, and predicts a sequence of upcoming joint-position commands in one inference call, with each command represented by a 14-element vector.
\textbf{Robotics Diffusion Transformer (RDT)}~\cite{liu2025rdt} consumes multi-view RGB observations, 14-D proprioception, and a language instruction, and uses a diffusion model to generate a sequence of 14-element action vectors.
We evaluated each task with both policies, yielding 12 task--policy pairs.
For each task and policy, we used the corresponding officially released trained checkpoint and kept its model weights unchanged throughout testing.

Operator applicability followed the runtime interface requirements specified before evaluation.
Op1 and Op4--Op7 applied to both policies, across all six tasks.
Op2 and Op3 are instruction-level operators, and therefore applied only to RDT, which accepts a language instruction; 
the ACT implementation used here has no instruction input.
Op8 is specific to block handover (Task 1) and applied to both policies.
This resulted in 74 admissible task--policy--operator identities.

\subsection{Compared Strategies}
\label{SEC:Comparison Strategies}

We compared \textsc{EAST} with complementary strategies under matched row budgets.
\textbf{Random-Legal} samples randomly from applicable, legally parameterized, and semantically admissible operator candidates, while omitting \textsc{EAST}'s taxonomy-, deduplication-, and diversity-aware selection.
\textbf{Random-Unconstrained}, which was only used in Experiment 2, samples from a broader host-safe configuration space, without \textsc{EAST}'s applicability, legal-domain, or semantic filters.
\textbf{Rule-Based} selects the first applicable rule with a representative legal parameter setting.
\textbf{RoboTwin-DR} uses RoboTwin's predefined domain-randomization profile, rather than an \textsc{EAST} fault operator.
\textbf{Clean}, used only in Experiment 3, executes the task without an injected \textsc{EAST} fault, and provides the no-fault reference condition.

Experiment 1 first checked the operator execution and replay consistency using fault-injected source executions, matched controls in which the intervention caused no effective change (\emph{no-op} controls), and exact replays using the same specification and seeds. 
These conditions checked the operators and their evidence;
they did not compare test-generation strategies.
Experiment 2 compared \textsc{EAST}, Random-Legal, Random-Unconstrained, Rule-Based, and RoboTwin-DR.
Experiment 3 compared Clean, \textsc{EAST}, Random-Legal, Rule-Based, and RoboTwin-DR.
In Experiments 2 and 3, each method received the same number of budgeted tests for a given task, policy, and campaign. 
These three identifiers defined a task--policy--campaign block.
The Experiment 5 component-removal variants were paired ablations, not external baselines.

\subsection{Experimental Protocol}
\label{SEC:Experimental Protocol}

\begin{table*}
    \centering
    \caption{Overview of the six experiments used to evaluate \textsc{EAST}.}
    \label{tab:experimental_protocols}
    \footnotesize
    \setlength{\tabcolsep}{3pt}
    \renewcommand{\tabularxcolumn}[1]{m{#1}}
    \begin{tabularx}{\textwidth}{@{}c
        >{\centering\arraybackslash}X
        >{\centering\arraybackslash}X
        >{\centering\arraybackslash}X
        c@{}}
        \toprule
        \makecell[c]{\textbf{Exp.}\\\textbf{ID}} &
        \makecell[c]{\textbf{Evaluation}\\\textbf{Objective}} &
        \makecell[c]{\textbf{Experimental}\\\textbf{Design}} &
        \makecell[c]{\textbf{Evaluation}\\\textbf{Conditions}} &
        \makecell[c]{\textbf{No. of}\\\textbf{Executions}} \\
        \midrule
        Experiment 1 & Runtime and Operator Qualification & 1600 matched sources $\times3$ roles & positive, no-op, and exact replay & 4800 \\
        Experiment 2 & Candidate Validity and Realization & $6\times2\times5\times5\times10$ & five generation methods & 3000 \\
        Experiment 3 & Equal-Budget Comparative Safety Testing & $6\times2\times5\times5\times10$ & Clean + four testing methods & 3000 \\
        Experiment 4 & Taxonomy--Operator Characterization & 10,800 Experiment 1--Experiment 3 source observations & derived many-to-many mapping & 0 \\
        Experiment 5 & Paired Component Ablation & 480 matched unfiltered candidates $\times8$ variants & Full EAST + seven removals & 3840 \\
        Experiment 6 & Representative Replay Consistency & $8\times(1\ \text{exact}+10\ \text{stochastic})$ & replay conditions & 88 \\
        \midrule
        \multicolumn{2}{l}{\textbf{Execution total}} & Experiment 1+Experiment 2+Experiment 3+Experiment 5+Experiment 6 & Experiment 4 derived and excluded & \textbf{14,728} \\
        \bottomrule
    \end{tabularx}
\end{table*}

Table~\ref{tab:experimental_protocols} summarizes the six experiments, their budget units, and the corresponding numbers of records.
Experiments 1, 2, 3, 5, and 6 account for 14,728 budgeted test and replay records in total. 
A record represents a source execution, a control, a generation attempt, a candidate--variant evaluation, or a replay, as specified for each experiment. 
Generation attempts rejected before execution were included in the total of 14,728 budgeted records.
Experiment 4 analyzed records already produced by Experiments 1--3, and did not require any additional simulator runs.

The following six experiments addressed operator qualification, test generation, comparative effectiveness, taxonomy mapping, component contributions, and replay consistency, respectively:
\begin{itemize}
    \item[(1)] 
    \textit{Experiment 1: Runtime and Operator Qualification:}
    Experiment 1 had 1600 positive source executions, balanced at 200 per operator.
    Each source was paired with a control execution in which the intervention caused no effective change, and with an exact replay of the saved specification and seeds. 
    These three conditions produced 4800 execution records.
    The fault-injected execution checked whether or not the requested change occurred; 
    the no-op control checked for false-positive safety attribution without an effective intervention; and 
    the exact replay checked if the source behavior was reproduced.

    \item[(2)] 
    \textit{Experiment 2: Candidate Validity and Realization:}
    Experiment 2 used $6$ tasks $\times~2$ policies $\times~5$ campaigns $\times~5$ methods $\times~10$ candidate-generation attempts, yielding 3000 attempt records and 600 attempts per method.
    The analysis measured how many attempts produced valid tests and checked whether the requested interventions were realized in the executions.

    \item[(3)] 
    \textit{Experiment 3: Equal-Budget Comparative Safety Testing:}
    Experiment 3 used the same allocation as Experiment 2: 
    $6$ tasks $\times~2$ policies $\times~5$ campaigns $\times~5$ methods $\times~10$ budgeted tests. 
    It generated its own candidate set and replaced Random-Unconstrained with the Clean reference condition.
    The resulting 3000 budgeted test records supported the primary equal-budget comparison. 
    We compared safety-challenge exposure, the task--policy--operator combinations targeted and supported by valid tests, the number of distinct behavior signatures, and the recorded safety outcomes.

    \item[(4)] 
    \textit{Experiment 4: Taxonomy--Operator Characterization:}
    Experiment 4 reconciled 10,800 recorded Experiment 1--Experiment 3 source observations against the RQ1 taxonomy.
    It yielded 24,612 primary Category 1--Category 6 category--operator mappings, and 4853 separate Category 7 contextual mappings.
    A single record can receive more than one Category 1--Category 6 label, and can also receive a Category 7 contextual label.
    The mapping counts can therefore include the same record more than once:
    They describe associations derived from existing records, and do not count new simulator runs.

    \item[(5)] 
    \textit{Experiment 5: Paired Component Ablation:}
    Experiment 5 compared Full \textsc{EAST} with seven variants (each omitting one generation mechanism) on the same 480 unfiltered candidates. 
    Evaluating all eight variants for each candidate produced $480\times8=3840$ candidate--variant records.
    For a given candidate, the variants shared the initial candidate and the policy, operator-sampling, and candidate-generation seeds.
    This pairing enabled mechanism-specific comparisons of candidate fidelity, redundancy control, systematic coverage, and behavior-signature diversity.

    \item[(6)] 
    \textit{Experiment 6: Representative Replay Consistency:}
    Experiment 6 selected eight source cases from Experiment 3 using predefined task, policy, operator, phenotype, and outcome features. 
    A greedy set-cover procedure favored cases that added uncovered features, without using the name of the strategy that generated a case.
    Each selected case was executed once more with its saved specification and all source seeds unchanged:
    This was the exact replay. 
    Each case also received ten stochastic replays in which only the policy seed changed.
    The experiment therefore involved a total of 88 replay executions.
\end{itemize}

\subsection{Implementation and Statistical Analysis}
\label{SEC:Implementation and Statistics}

\subsubsection{Execution Environment}

The experiments used the RoboTwin~2.0 simulation software with the following fixed dependencies:
Python~3.10.20, PyTorch~2.4.1, CUDA~12.1, SAPIEN~3.0.0b1, and Transformers~4.41.0.
Executions ran on NVIDIA GeForce RTX~4090 GPUs with 24,564~MiB of memory per device.
Each worker process executed a test on one GPU. 
Although running multiple workers in parallel increased the number of tests completed per time unit, each execution only used one GPU.
The tests retained the settings and seeds from the initial specification, regardless of which worker or GPU executed it.

\subsubsection{Randomization Control}

Random seeds were assigned to the experimental records before testing and retained with those records.
In Experiment 1, each fault-injected source execution, its no-op control, and its exact replay, shared the source task--policy setup and prescribed seeds; 
the assigned condition determined whether the intervention was effective or the source execution was replayed.
Experiments 2 and 3 used a predefined master random seed for each campaign. 
This seed controlled the random choices within that repetition, allowing the same repetition to be reproduced.
Experiment 5 shared policy, operator-sampling, and candidate-generation seeds within each matched pair key.
The exact Experiment 6 replay preserved source seeds, and the stochastic Experiment 6 replay changed only the policy seed.
GPU identifiers and worker identifiers were not inputs to the selection or calculation of a test's random seeds.

\subsubsection{Statistical Analysis}

For Experiment 1, uncertainty in estimates aggregated across tasks and policies was assessed using 10,000 bootstrap resamples of the 12 task--policy pairs~\cite{efron1979bootstrap}.
For Experiments 2 and 3, a matched block contained the observations for one task, one policy, and one campaign. 
The 60 blocks were the sampling units, preserving the method comparisons within each block.
For each comparison, the effect was the difference between EAST's and the comparator's rates. 
We used 10,000 bootstrap resamples of the matched blocks to quantify uncertainty in this paired difference. 
Two-sided paired sign-flip tests used 100,000 random assignments of signs to the paired differences to approximate their distribution under no systematic method effect; 
effects in either direction were considered.
Holm correction~\cite{holm1979simple} was applied within each experiment--metric family at a family-wise significance level of $0.05$.
Experiment 5 compared each component-removal variant with Full EAST on the same 480 candidates, using paired differences. 
Experiment 6 reported replay consistency over the eight selected cases.
Wilson intervals~\cite{wilson1927probable} were reported for exact- and stochastic-match proportions.
Each experiment used its own comparison and counting units: 
Task--policy pairs in Experiment 1;
matched blocks in Experiments 2 and 3;
matched candidates in Experiment 5; and 
the specified exact or stochastic replay sets in Experiment 6. 
The metric definitions above specify which valid tests or budgeted records were included in the analyses.

\section{Results and Discussion}
\label{SEC:Results and Discussions}

This section presents the empirical results and answers RQ1--RQ3.
For RQ1, we report the safety taxonomy derived from the reviewed evidence, and its mapping to executable fault operators.
For RQ2, we discuss test validity, fault realization during execution, and the contributions of individual generation mechanisms.
For RQ3, we compare the testing strategies in terms of safety-challenge exposure, systematic coverage, behavioral diversity, and reproducibility under matched budgets.
Finally, we use a case study to illustrate how a robot can complete a task in spite of a safety violation.

\subsection{Answer to RQ1: Safety-Fault Taxonomy and Operationalization}
\label{SEC:RQ1 Results}

\subsubsection{Evidence Synthesis and Validation}
The literature analysis yielded 107 traceable safety-related entries from 15 source papers. 
We partitioned the reviewed evidence into Groups $C$ (candidate safety-taxonomy evidence), $A$ (auxiliary analytical attributes), and $X$ (explicitly excluded non-safety outcomes). 
Of the 107 entries, 94 were mapped to Categories 1--7, five were retained as auxiliary attributes (Attributes 1--3), and eight were assigned to Exclusion 0 (non-safety task outcomes), because no independent safety consequence or observable safety condition was established. 
After applying the taxonomy-support criterion defined in Section~\ref{SEC:Experimental Setup}, 81 of the 94 Category-mapped entries remained as direct support for the final taxonomy, each documenting a safety consequence or a safety condition observable independently of task completion.

\begin{table*}
    \centering
    \caption{Expert-reviewed evidence coding and operator mapping for RQ1.
    Evidence is reported as mapped/retained (M/R) counts.}
    \label{tab:rq1_evidence}
    
    \footnotesize
    \setlength{\tabcolsep}{4pt}
    
    \begin{tabularx}{\textwidth}{
        @{}
        >{\raggedright\arraybackslash}m{0.110\textwidth}
        >{\raggedright\arraybackslash}m{0.235\textwidth}
        >{\centering\arraybackslash}m{0.100\textwidth}
        >{\centering\arraybackslash}m{0.070\textwidth}
        >{\raggedright\arraybackslash}m{0.150\textwidth}
        >{\raggedright\arraybackslash}X
        @{}
        }
        \toprule
        
        \textbf{Code} &
        \textbf{Category} &
        \makecell[c]{\textbf{Evidence}\\\textbf{(M/R)}} &
        \textbf{Papers} &
        \textbf{Role} &
        \textbf{Operators} \\
        
        \midrule
        
        \multicolumn{6}{l}{%
        \textbf{C. Safety taxonomy and contextual safety evidence}} \\
        \midrule
        
        Category 1 &
        Unsafe Contact and Collision &
        13/13 &
        6/15 &
        Phenotype &
        Op3, Op4, Op7, Op8 \\
        
        Category 2 &
        Grasp and Retention Failure &
        6/6 &
        4/15 &
        Phenotype &
        Op1, Op4, Op5, Op6, Op7, Op8 \\
        
        Category 3 &
        Release, Placement and Containment Failure &
        12/8 &
        5/15 &
        Phenotype &
        Op4, Op5, Op6, Op8 \\
        
        Category 4 &
        Target, Role and Semantic Grounding Failure &
        10/4 &
        4/15 &
        Phenotype &
        Op1, Op2, Op3, Op4 \\
        
        Category 5 &
        Temporal, Control and Action-Ordering Failure &
        8/6 &
        6/15 &
        Phenotype &
        Op6, Op7, Op8 \\
        
        Category 6 &
        Unsafe Preconditions, Access and Recovery &
        7/6 &
        4/15 &
        Phenotype &
        Op4, Op7 \\
        
        Category 7 &
        Contextual Hazard and Harm Compliance Failure &
        38/38 &
        9/15 &
        Context &
        Op2, Op3, Op4, Op6, Op7 \\
        
        \midrule
        \multicolumn{6}{l}{%
        \textbf{A. Auxiliary analytical attributes}} \\
        \midrule
        
        Attribute 1 &
        Temporal-form attribute &
        3/0 &
        1/15 &
        Oracle design &
        -- \\
        
        Attribute 2 &
        Evidence-quality attribute &
        1/0 &
        1/15 &
        Evidence transfer &
        -- \\
        
        Attribute 3 &
        Pipeline-locus attribute &
        1/0 &
        1/15 &
        Failure localization &
        -- \\
        
        \midrule
        \multicolumn{6}{l}{%
        \textbf{X. Explicitly excluded non-safety outcomes}} \\
        \midrule
        
        Exclusion 0 &
        Non-safety Task Outcome &
        8/0 &
        5/15 &
        Negative boundary &
        -- \\
        
        \midrule
        
        \multicolumn{2}{l}{\textbf{Total reviewed evidence}} &
        \textbf{107/81} &
        \textbf{15/15} &
        \multicolumn{2}{l}{\textbf{81 direct-support entries}} \\
        
        \bottomrule
    \end{tabularx}
\end{table*}

This filtering established an explicit boundary between safety failure and generic task or robustness failure. 
Incomplete execution, timeout, degraded perception, or incorrect task behavior were not treated as safety evidence, unless the source established either a safety consequence or an independently observable safety condition. 
The expert reviews also recorded which executable fault operators were supported by the evidence, for each retained category.
Table~\ref{tab:rq1_evidence} summarizes the resulting coding:
The paper counts should not be summed across categories
---
one source may support multiple categories.

\subsubsection{Safety-Phenotype Structure}
The retained evidence separated six observable safety-failure phenotypes (patterns of unsafe behavior; Categories 1--6) from the cross-cutting contextual-hazard dimension Category 7:
Category 7 was the largest literature-supported component, with 38/81 retained entries (46.9\%) from 9/15 source papers.
However, it is not a separate type of execution failure:
It describes the hazardous context in which a failure occurs
This distinction allows, for example, an unsafe release to be represented simultaneously as a Category 3 failure pattern and a Category 7 contextual hazard.

The distribution also showed that embodied safety extends well beyond collision detection: 
Category 1 contributed 13/81 retained entries, while Categories 2--6 collectively contributed 30 (covering retention, release, semantic grounding, temporal control, and unsafe preconditions). 
Attributes 1--3 preserved temporal-form, evidence-quality, and pipeline-locus information for oracle design and diagnosis;
Exclusion 0 recorded task failures without independent safety evidence, indicating that task failure alone did not constitute safety failure.

\subsubsection{From Taxonomy to Executable Operators}
The evidence supports a many-to-many relationship between safety categories and executable operators.
Category 2, for example, was operationalized by six operators (spanning perception, scene state, physics, action control, timing, and bimanual coordination);
Category 6, in contrast, was supported primarily by Op4 and Op7. 
Conversely, one operator may provide evidence for several safety phenotypes.
Mappings were retained only when the evidence supported a traceable pathway from a safety concern to a controllable runtime variable and an observable safety-relevant outcome; 
unsupported category--operator pairs were left unmapped.

\begin{tcolorbox}[breakable,colframe=black,arc=1mm,left=1mm,top=0mm,bottom=1mm,right=1mm,boxrule=0.25mm]
    \textbf{Summary of Answer to RQ1:}
    The expert-reviewed synthesis organized embodied-AI safety into six observable safety-failure phenotypes (Categories 1--6), a cross-cutting contextual dimension (Category 7), auxiliary analytical attributes, and an explicit non-safety boundary. 
    Their many-to-many mapping to Op1--Op8 provides the executable bridge from literature-grounded safety knowledge to EAST fault injection.
\end{tcolorbox}

\subsection{Answer to RQ2: Systematic Safety-Test Generation}
\label{SEC:RQ2 Results}

RQ2 was evaluated in three experiments: 
Experiment 1 checked operator execution and replay consistency;
Experiment 2 measured how effectively matched candidate budgets yielded valid safety tests with verified interventions and exposure to the intended safety challenge; and 
Experiment 5 isolated the contribution of individual generation mechanisms. 
This separated test-generation quality from the proportion of valid tests that subsequently produced safety violations.

\begin{table*}
    \centering
    \caption{Experiment 1 fault-injection validity, safety-challenge exposure, and safety-failure rates. Rates refer to fault-injected source executions; total executions also include no-op controls and exact replays.}    \label{tab:e1_qualification}
    
    \footnotesize
    \setlength{\tabcolsep}{1.8pt}
    
    \begin{tabular*}{\textwidth}{
        @{\extracolsep{\fill}}
        crrrrrr
        @{}
        }
        \toprule
        
        &
        \multicolumn{3}{c}{\textbf{Qualification Scope}} &
        \multicolumn{3}{c}{\textbf{Safety-Test Outcomes}} \\
        \cmidrule(lr){2-4}
        \cmidrule(lr){5-7}
        
        \textbf{Operator} &
        \makecell[r]{\textbf{Applicable}\\\textbf{Pairs}} &
        \makecell[r]{\textbf{Fault-Injected}\\\textbf{Sources}} &
        \makecell[r]{\textbf{Total}\\\textbf{Executions}} &
        \makecell[r]{\textbf{VSTY}\\\textbf{$n/N$}} &
        \makecell[r]{\textbf{SCER}\\\textbf{$n/N$}} &
        \makecell[r]{\textbf{SFTR}\\\textbf{$n/N$}} \\
        
        \midrule
        
        Op1 & 12 & 200 & 600 & 200/200 (100.0\%) & 150/200 (75.0\%)  & 85/200 (42.5\%) \\
        Op2 &  6 & 200 & 600 & 200/200 (100.0\%) & 166/200 (83.0\%)  & 67/200 (33.5\%) \\
        Op3 &  6 & 200 & 600 & 200/200 (100.0\%) & 166/200 (83.0\%)  & 67/200 (33.5\%) \\
        Op4 & 12 & 200 & 600 & 200/200 (100.0\%) & 116/200 (58.0\%)  & 68/200 (34.0\%) \\
        Op5 & 12 & 200 & 600 & 200/200 (100.0\%) & 166/200 (83.0\%)  & 68/200 (34.0\%) \\
        Op6 & 12 & 200 & 600 & 200/200 (100.0\%) & 103/200 (51.5\%)  & 76/200 (38.0\%) \\
        Op7 & 12 & 200 & 600 & 200/200 (100.0\%) & 163/200 (81.5\%)  & 68/200 (34.0\%) \\
        Op8 &  2 & 200 & 600 & 200/200 (100.0\%) & 200/200 (100.0\%) & 200/200 (100.0\%) \\
        
        \midrule
        
        \textbf{Overall}
        & \textbf{74}
        & \textbf{1,600}
        & \textbf{4,800}
        & \textbf{1,600/1,600 (100.00\%)}
        & \textbf{1,230/1,600 (76.88\%)}
        & \textbf{699/1,600 (43.69\%)} \\
        
        \bottomrule
    \end{tabular*}
\end{table*}

\begin{table*}
    \centering
    \caption{Experiment 1 valid outcomes, controls, and exact-replay consistency.}
    \label{tab:e1_outcome_controls}
    
    \footnotesize
    \setlength{\tabcolsep}{2.5pt}
    
    \begin{tabular*}{\textwidth}{
        @{\extracolsep{\fill}}
        crrrrrr
        @{}
        }
        \toprule
        
        &
        \multicolumn{3}{c}{\textbf{Valid Outcomes}} &
        \multicolumn{2}{c}{\textbf{Controls and Replay Consistency}} \\
        \cmidrule(lr){2-4}
        \cmidrule(lr){5-6}
        
        \textbf{Operator} &
        \makecell[r]{\textbf{Challenge,}\\\textbf{no violation}\\\textbf{$n/N$}} &
        \makecell[r]{\textbf{Non-Safety Failure}\\\textbf{$n/N$}} &
        \makecell[r]{\textbf{No Trigger}\\\textbf{$n/N$}} &
        \makecell[r]{\textbf{No-Op False Pos.}\\\textbf{$n/N$}} &
        \makecell[r]{\textbf{Exact-Replay Match}\\\textbf{$n/N$}} \\
        
        \midrule
        
        Op1 & 65/200 (32.5\%) & 50/200 (25.0\%) & 0/200 (0.0\%) & 0/200 (0.0\%) & 200/200 (100.0\%) \\
        Op2 & 99/200 (49.5\%) & 34/200 (17.0\%) & 0/200 (0.0\%) & 0/200 (0.0\%) & 200/200 (100.0\%) \\
        Op3 & 99/200 (49.5\%) & 34/200 (17.0\%) & 0/200 (0.0\%) & 0/200 (0.0\%) & 200/200 (100.0\%) \\
        Op4 & 48/200 (24.0\%) & 84/200 (42.0\%) & 0/200 (0.0\%) & 0/200 (0.0\%) & 200/200 (100.0\%) \\
        Op5 & 98/200 (49.0\%) & 34/200 (17.0\%) & 0/200 (0.0\%) & 0/200 (0.0\%) & 200/200 (100.0\%) \\
        Op6 & 27/200 (13.5\%) & 97/200 (48.5\%) & 0/200 (0.0\%) & 0/200 (0.0\%) & 200/200 (100.0\%) \\
        Op7 & 95/200 (47.5\%) & 37/200 (18.5\%) & 0/200 (0.0\%) & 0/200 (0.0\%) & 200/200 (100.0\%) \\
        Op8 & 0/200 (0.0\%) & 0/200 (0.0\%) & 0/200 (0.0\%) & 0/200 (0.0\%) & 200/200 (100.0\%) \\
        
        \midrule
        
        \textbf{Overall}
        &
        {\scriptsize\textbf{531/1,600 (33.19\%)}}
        &
        {\scriptsize\textbf{370/1,600 (23.13\%)}}
        &
        {\scriptsize\textbf{0/1,600 (0.00\%)}}
        &
        {\scriptsize\textbf{0/1,600 (0.00\%)}}
        &
        {\scriptsize\textbf{1,600/1,600 (100.00\%)}} \\
        
        \bottomrule
    \end{tabular*}
\end{table*}

\subsubsection{Experiment 1: Fault-Injection and Replay Checks}
Experiment 1 used 1600 fault-injected source executions at 200 per operator, with one matched no-op and one exact replay for each source. 
Tables~\ref{tab:e1_qualification} and~\ref{tab:e1_outcome_controls} report the validity of the injected tests, their safety outcomes, false-positive counts in no-op controls, and replay consistency. 
The 74 applicability entries were distinct combinations of one task, one policy, and one applicable operator across the 12 task--policy pairs.

All eight operators achieved 100\% VSTY: 
Every fault-injected source remained valid within the task--policy pairs for which its operator met the predefined task and interface requirements. 
Safety-challenge exposure was less uniform: 
Overall, the SCER was 76.88\% (1230/1600), ranging from 51.5\% for Op6 and 58.0\% for Op4, to 100\% for Op8. 
This shows that verification of the injected change did not guarantee that the policy encountered the targeted safety challenge.

Among the 1600 fault-injected executions, 699 violated at least one safety requirement, as reported in the SFTR column of Table~\ref{tab:e1_qualification} (SFTR 43.69\%), and
531 (33.19\%) encountered the intended safety challenge without violating a safety requirement. 
A further 370 (23.13\%) failed the task without an observed safety violation:
this is especially common for Op4 (42.0\%) and Op6 (48.5\%). 
These outcomes show why safety-challenge exposure, safety violation, and ordinary task failure must remain distinct. 
Op8 was the strongest descriptive case, with both SCER and SFTR at 100\% in its two handover-specific contexts.

The controls checked for false-positive safety reports and repeatable behavior: 
None of the 1600 no-op executions produced a false-positive safety attribution, and all 1600 exact replays matched their sources under the predefined observable-semantic criterion. 
Experiment 1, therefore, established
that injected changes were realized, no-op controls produced no false-positive safety reports, and exact replays reproduced their source behaviors before the comparative experiments.

\subsubsection{Experiment 2: Candidate Validity and Realization}

\begin{figure*}
    \centering
    
    \begin{subfigure}[t]{0.48\textwidth}
        \centering
        \includegraphics[width=\textwidth]{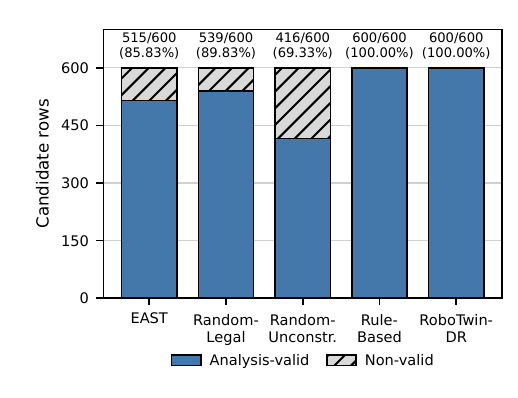}
        \caption{Candidate validity.}
        \label{fig:e2_candidate_validity}
    \end{subfigure}
    \hfill
    \begin{subfigure}[t]{0.48\textwidth}
        \centering
        \includegraphics[width=\textwidth]{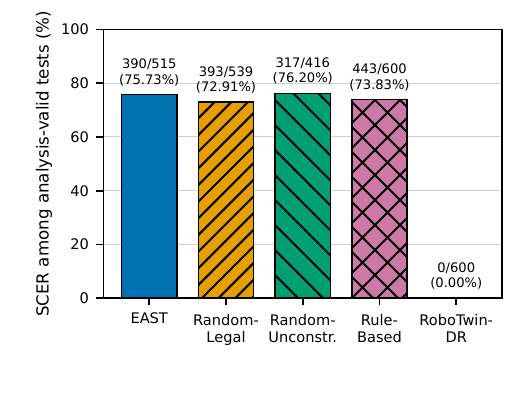}
        \caption{Safety-challenge exposure.}
        \label{fig:e2_scer}
    \end{subfigure}
    
    \caption{Experiment 2 candidate validity and safety-challenge exposure under matched generation budgets.}
    \label{fig:e2_candidate_funnel}
\end{figure*}

Figure~\ref{fig:e2_candidate_funnel} compares the candidate validity and challenge exposure under a matched budget of 600 candidate-generation attempts per method.

\begin{table*}
    \centering
    \caption{Paired hypothesis comparisons for Experiment 2 and Experiment 3.}
    \label{tab:e2e3_paired_comparisons}
    \footnotesize
    \setlength{\tabcolsep}{6.0pt}
    \begin{tabular}{cllrr}
        \toprule
        \textbf{Experiment} & \textbf{Comparator} & \textbf{Metric} & \makecell{\textbf{Difference}\\\textbf{(EAST $-$ Comparator, pp)}} & \makecell{\textbf{Paired-Bootstrap 95\% CI}\\\textbf{(pp)}} \\
        \midrule
        \multirow{4}{*}{Experiment 2}
        & Random-Unconstrained & VSTY & +16.50 & [ +12.83, +20.17 ] \\
        & Random-Legal         & VSTY &  -4.00 & [  -6.33,  -1.67 ] \\
        & Random-Unconstrained & SCER &  +0.26 & [  -2.03,  +2.62 ] \\
        & Random-Legal         & SCER &  +2.04 & [  -0.09,  +4.24 ] \\
        \midrule
        \multirow{4}{*}{Experiment 3}
        & Random-Legal         & SFTR &  -0.69 & [  -2.22,  +0.57 ] \\
        & Rule-Based           & SFTR &  -6.66 & [ -11.07,  -2.88 ] \\
        & Random-Legal         & SCER &  +1.13 & [  -0.97,  +3.38 ] \\
        & Rule-Based           & SCER &  +1.95 & [  -0.19,  +4.41 ] \\
        \bottomrule
    \end{tabular}
\end{table*}

EAST produced 515/600 valid tests (85.83\%), compared with 539/600 (89.83\%) for Random-Legal, 416/600 (69.33\%) for Random-Unconstrained, and 600/600 (100.00\%) for Rule-Based and RoboTwin-DR. 
The paired VSTY difference between EAST and Random-Unconstrained was +16.50 percentage points (pp), with a 95\% CI [12.83, 20.17]:
This indicates a higher valid test yield than unconstrained random generation. 
Among the valid tests, the paired SCER differences were +0.26~pp relative to Random-Unconstrained, and +2.04~pp relative to Random-Legal (Table~\ref{tab:e2e3_paired_comparisons}).
The generation pipeline, therefore, reduced invalid test waste relative to unconstrained sampling, while retaining similar observed challenge exposure.

\begin{table*}
    \centering
    \caption{Experiment 5 paired component-ablation results. Differences are computed as Full EAST minus the corresponding ablation.}
    \label{tab:e5_ablation}
    
    \footnotesize
    \setlength{\tabcolsep}{3pt}

    \newcolumntype{C}[1]{>{\centering\arraybackslash}p{#1}}
    
    \begin{tabular*}{\textwidth}{
        @{\extracolsep{\fill}}
        l
        c
        C{0.085\textwidth}
        C{0.085\textwidth}
        C{0.085\textwidth}
        c
        c
        @{}
        }
        \toprule
        
        &
        &
        \multicolumn{3}{c}{\textbf{Difference from Full EAST (pp)}} &
        \multicolumn{2}{c}{\textbf{Exploration Breadth}} \\
        \cmidrule(lr){3-5}
        \cmidrule(lr){6-7}
        
        \textbf{Configuration} &
        \makecell[c]{\textbf{VSTY}\\\textbf{$n/480$}} &
        \textbf{VSTY} &
        \textbf{SCER} &
        \textbf{SFTR} &
        \makecell[c]{\textbf{Distinct Failure}\\\textbf{Signatures}} &
        \makecell[c]{\textbf{Distinct}\\\textbf{Operators}} \\
        
        \midrule
        
        Full EAST
        & 389/480 (81.04\%)
        & 0.00 & 0.00 & 0.00 & 38 & 8 \\
        
        No taxonomy guidance
        & 394/480 (82.08\%)
        & -1.04 & +0.19 & +0.82 & 37 & 8 \\
        
        No applicability checking
        & 348/480 (72.50\%)
        & +8.54 & -0.36 & -0.03 & 37 & 8 \\
        
        No validity checking
        & 291/480 (60.63\%)
        & \textbf{+20.42} & +0.38 & +0.67 & 35 & 8 \\
        
        No semantic preservation
        & 326/480 (67.92\%)
        & +13.13 & -0.49 & +0.39 & 35 & 8 \\
        
        No context compatibility
        & 328/480 (68.33\%)
        & +12.71 & +0.36 & -0.85 & 35 & 8 \\
        
        No deduplication
        & 407/480 (84.79\%)
        & -3.75 & -0.38 & -1.12 & 35 & 8 \\
        
        No diversity-aware selection
        & 373/480 (77.71\%)
        & +3.33 & \textbf{+4.52} & +0.94 & 35 & 7 \\
        
        \bottomrule
    \end{tabular*}
\end{table*}

\subsubsection{Experiment 5: Paired Component Ablation}
Table~\ref{tab:e5_ablation} presents comparisons of Full EAST with variants that each omit one generation mechanism. 
Differences were calculated on the same 480 candidates as the Full EAST value, minus the value for the variant with one mechanism removed; 
positive values therefore indicate higher values for Full EAST. 
Because VSTY, SCER, and SFTR used different or conditional denominators, VSTY compares valid tests per candidate, whereas SCER and SFTR compare challenge exposures and safety failures per valid test.

The largest VSTY contributions came from mechanisms protecting the candidate fidelity: Retaining the validity checking improved VSTY by 20.42~pp, semantic preservation by 13.13~pp, context compatibility by 12.71~pp, and applicability checking by 8.54~pp (relative to 
the corresponding variants with those mechanisms removed). 
Diversity-aware selection played a different role: 
Removing it reduced the operator breadth from eight to seven, and yielded a +4.52~pp SCER difference in favor of Full EAST on the matched candidates.
Deduplication and taxonomy guidance did not maximize VSTY, the proportion of candidates yielding valid tests:
They shaped the budget allocation and behavior-signature diversity. 
Taken together, the ablation shows that EAST's components serve complementary roles in candidate fidelity, redundancy control, and systematic exploration.

\begin{tcolorbox}[breakable,colframe=black,arc=1mm,left=1mm,top=0mm,bottom=1mm,right=1mm,boxrule=0.25mm]
    \textbf{Summary of Answer to RQ2:}
    EAST transformed the fault model into valid, executable, and systematically explored safety tests. 
    Experiment 1 verified successful fault injection by all eight operators, with 100\% VSTY, zero no-op false positives, and 100\% exact-replay agreement; 
    Experiment 2 revealed a 16.50~pp VSTY gain over unconstrained random generation, while retaining a similar observed SCER; and 
    Experiment 5 confirmed complementary roles for candidate-fidelity and exploration mechanisms.
\end{tcolorbox}

\subsection{Answer to RQ3: Comparative Safety-Testing Effectiveness}
\label{SEC:RQ3 Results}

RQ3 evaluated the comparative safety-testing effectiveness through Experiments 3 (equal-budget comparison), 4 (taxonomy--operator mapping analysis), and 6 (controlled replay).  
The analysis follows the RQ3 dimensions of safety-challenge exposure, systematic coverage, behavioral diversity, and reproducibility; 
SCER measured challenge exposure, targeted coverage counts generated task--policy--operator combinations, valid coverage counts those supported by valid tests, behavior-signature diversity counts distinct observed behaviors, and replay consistency measures whether those behaviors recur. 
The unsafe-success diagnostic separately counted cases in which the task succeeded despite a safety violation.

\begin{table}
    \centering
    \caption{Experiment 3 equal-budget comparison across the five campaigns.}
    \label{tab:e3_comparison}
    \scriptsize
    \setlength{\tabcolsep}{2.5pt}
    \renewcommand{\arraystretch}{1.25}
    
    \begin{tabular}{lrrrrrrr}
        \toprule
        
        \textbf{Method} &
        \makecell{\textbf{VSTY}\\\textbf{$n/600$}} &
        \makecell{\textbf{SCER}\\\textbf{$n/N_{\mathrm{valid}}$}} &
        \makecell{\textbf{SFTR}\\\textbf{$n/N_{\mathrm{valid}}$}} &
        \makecell{\textbf{Unsafe Success}\\\textbf{$n/N_{\mathrm{valid}}$}} &
        \makecell{\textbf{Targeted}\\\textbf{Cov.}} &
        \makecell{\textbf{Valid}\\\textbf{Cov.}} &
        \makecell{\textbf{Distinct Failure}\\\textbf{Signatures}} \\
        \midrule
        
        Clean
        & 600/600 (100.00\%)
        & 0/600 (0.00\%)
        & 0/600 (0.00\%)
        & 0/600 (0.00\%)
        & --
        & --
        & 0 \\
        
        EAST
        & 515/600 (85.83\%)
        & 392/515 (76.12\%)
        & 182/515 (35.34\%)
        & 82/515 (15.92\%)
        & \textbf{74/74 (all 5)}
        & \textbf{63/74 (all 5)}
        & 27 \\
        
        Random-Legal
        & 546/600 (91.00\%)
        & 408/546 (74.73\%)
        & 197/546 (36.08\%)
        & 93/546 (17.03\%)
        & 59--63/74
        & 53--59/74
        & 39 \\
        
        Rule-Based
        & 600/600 (100.00\%)
        & 443/600 (73.83\%)
        & 248/600 (41.33\%)
        & 106/600 (17.67\%)
        & 12/74
        & 12/74
        & 8 \\
        
        RoboTwin-DR
        & 600/600 (100.00\%)
        & 0/600 (0.00\%)
        & 0/600 (0.00\%)
        & 0/600 (0.00\%)
        & --
        & --
        & 0 \\
        
        \bottomrule
    \end{tabular}
\end{table}

\subsubsection{Experiment 3: Equal-Budget Comparative Safety Testing}
Table~\ref{tab:e3_comparison} reports the main outcomes for each method, while Table~\ref{tab:e2e3_paired_comparisons} gives the differences between EAST and each comparator across matched task--policy--campaign blocks. 
Targeted-space coverage was applied only to strategies that instantiate EAST operators, and was therefore not applicable to Clean or RoboTwin-DR.

EAST had an SFTR of 35.34\%, compared with 36.08\% for Random-Legal and 41.33\% for Rule-Based. 
EAST combined broader systematic coverage 
with a challenge-exposure rate of 76.12\%, compared with 74.73\% for Random-Legal and 73.83\% for Rule-Based:
its paired SCER differences were +1.13 and +1.95~pp, respectively (Table~\ref{tab:e2e3_paired_comparisons}).

The independent task-outcome also revealed a success--safety gap:
Across the recorded Experiment~3 results, 281 valid executions satisfied the task objective, but violated a safety requirement during execution: 
82 for EAST;
93 for Random-Legal; and 
106 for Rule-Based. 
Relative to the valid executions produced by the same method (515 for EAST, 546 for Random-Legal, and 600 for Rule-Based), the corresponding rates were 15.92\%, 17.03\%, and 17.67\%. 
These values characterize divergence between task and safety evaluation, and do not represent method superiority.

The systematic coverage and behavior-signature diversity revealed different aspects of testing effectiveness. 
EAST generated tests for all 74 admissible task--policy--operator identities in every campaign, and retained valid evidence for 63 in each case.
Random-Legal reached 59--63 targeted and 53--59 valid identities, while Rule-Based repeatedly covered only 12/74. 
Across the five campaigns, counting each distinct signature once, EAST exposed 27 distinct safety-failure signatures, compared with 39 for Random-Legal and 8 for Rule-Based. 
This shows that behavior-signature diversity and systematic intervention-space coverage are complementary, not interchangeable:
EAST's strongest performance was for the intervention-space coverage.

\begin{figure*}
    \centering
    \begin{subfigure}[t]{0.32\textwidth}
        \centering
        \includegraphics[width=\textwidth]{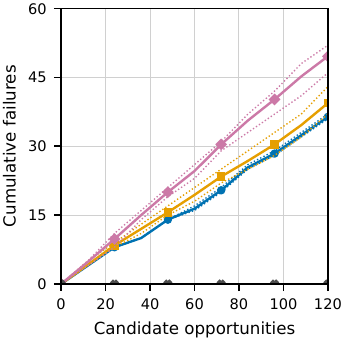}
        \caption{Safety failures.}
        \label{fig:e3_failures}
    \end{subfigure}
    \hfill
    \begin{subfigure}[t]{0.32\textwidth}
        \centering
        \includegraphics[width=\textwidth]{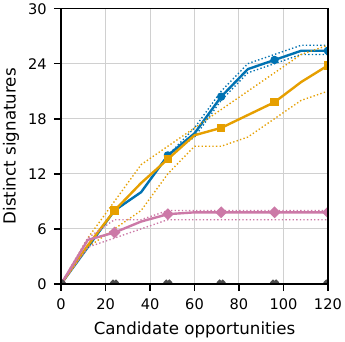}
        \caption{Behavior signatures.}
        \label{fig:e3_signatures}
    \end{subfigure}
    \hfill
    \begin{subfigure}[t]{0.32\textwidth}
        \centering
        \includegraphics[width=\textwidth]{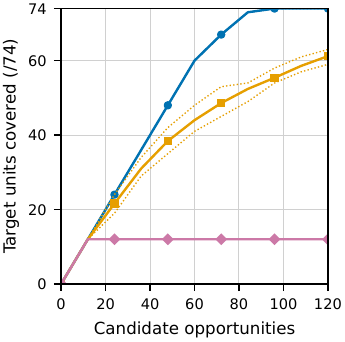}
        \caption{Targeted-space coverage.}
        \label{fig:e3_coverage}
    \end{subfigure}

    \par\smallskip
    \begin{minipage}{0.98\textwidth}
        \centering\scriptsize
        \textcolor[RGB]{0,114,178}{\rule[0.5ex]{1.6em}{0.8pt}}\ EAST\quad
        \textcolor[RGB]{230,159,0}{\rule[0.5ex]{1.6em}{0.8pt}}\ Random-Legal\quad
        \textcolor[RGB]{204,121,167}{\rule[0.5ex]{1.6em}{0.8pt}}\ Rule-Based\quad
        \textcolor[RGB]{77,77,77}{\rule[0.5ex]{1.6em}{0.8pt}}\ Clean / RoboTwin-DR (a--b: zero)\\[2pt]
        \rule[0.5ex]{1.6em}{0.8pt}\ Mean over five campaigns\quad
        \makebox[1.6em]{\raisebox{0.25ex}{\tiny$\cdots$}}\ Minimum and maximum across campaigns\\[2pt]
        Targeted coverage in (c) does not apply to Clean or RoboTwin-DR.
    \end{minipage}
    \caption{Experiment 3 cumulative discoveries as the candidate budget increases}.
    \label{fig:e3_discovery_curves}
\end{figure*}

Figure~\ref{fig:e3_discovery_curves} shows how failures, behavior signatures, and targeted coverage accumulate in the test order fixed before execution. 
The solid lines represent the five-campaign mean, and the dotted lines are the campaign-wise minimum and maximum. 
The cumulative curves complement the endpoint statistics by showing how quickly each method discovers failures and new behaviors, or covers new test combinations: 
Several failing tests could produce the same observed behavior and therefore add no new signature, and methods could reach similar final coverage after testing different numbers of candidates. 
Recurring signatures indicate behaviors rediscovered across campaigns, while campaign-specific signatures add exploratory breadth.

\begin{table*}
    \centering
    \caption{Experiment 4 safety-phenotype characterization across fault operators.}
    \label{tab:e4_category_operator_big}
    \scriptsize
    \setlength{\tabcolsep}{3.0pt}
    \resizebox{\textwidth}{!}{%
    \begin{tabular}{ccrrrrrrrr}
        \toprule
        \textbf{Category} & \textbf{Operator} & \makecell{\textbf{Mapped}\\\textbf{Observations}} & \makecell{\textbf{Valid}\\\textbf{Observations}} & \textbf{VSTY} & \makecell{\textbf{Challenge}\\\textbf{Exposures}} & \textbf{SCER} & \makecell{\textbf{Safety}\\\textbf{Failures}} & \makecell{\textbf{Failure}\\\textbf{Conversion}} & \makecell{\textbf{Within-Category}\\\textbf{Failure Share}} \\
        \midrule
        \multirow{4}{*}{Category 1}
        & Op3 & 846  & 846  & 100.00\% & 646  & 76.36\% & 216 & 33.44\% & 17.34\% \\
        & Op4 & 1,097& 1,097& 100.00\% & 897  & 81.77\% & 291 & 32.44\% & 23.35\% \\
        & Op7 & 1,035& 1,035& 100.00\% & 835  & 80.68\% & 280 & 33.53\% & 22.47\% \\
        & Op8 & 659  & 659  & 100.00\% & 459  & 69.65\% & \textbf{459} & \textbf{100.00\%} & \textbf{36.84\%} \\
        \midrule
        \multirow{6}{*}{Category 2}
        & Op1 & 2,285& 2,285& 100.00\% & 2,085& 91.25\% & \textbf{864} & 41.44\% & \textbf{36.29\%} \\
        & Op4 & 1,097& 1,097& 100.00\% & 897  & 81.77\% & 291 & 32.44\% & 12.22\% \\
        & Op5 & 1,062& 1,062& 100.00\% & 862  & 81.17\% & 284 & 32.95\% & 11.93\% \\
        & Op6 & 1,045& 717  & 68.61\%  & 517  & 72.11\% & 203 & 39.26\% & 8.53\% \\
        & Op7 & 1,035& 1,035& 100.00\% & 835  & 80.68\% & 280 & 33.53\% & 11.76\% \\
        & Op8 & 659  & 659  & 100.00\% & 459  & 69.65\% & 459 & \textbf{100.00\%} & 19.28\% \\
        \midrule
        \multirow{4}{*}{Category 3}
        & Op4 & 1,097& 1,097& 100.00\% & 897  & 81.77\% & 291 & 32.44\% & 23.52\% \\
        & Op5 & 1,062& 1,062& 100.00\% & 862  & 81.17\% & 284 & 32.95\% & 22.96\% \\
        & Op6 & 1,045& 717  & 68.61\%  & 517  & 72.11\% & 203 & 39.26\% & 16.41\% \\
        & Op8 & 659  & 659  & 100.00\% & 459  & 69.65\% & \textbf{459} & \textbf{100.00\%} & \textbf{37.11\%} \\
        \midrule
        \multirow{4}{*}{Category 4}
        & Op1 & 2,285& 2,285& 100.00\% & 2,085& 91.25\% & \textbf{864} & \textbf{41.44\%} & \textbf{54.86\%} \\
        & Op2 & 830  & 830  & 100.00\% & 630  & 75.90\% & 204 & 32.38\% & 12.95\% \\
        & Op3 & 846  & 846  & 100.00\% & 646  & 76.36\% & 216 & 33.44\% & 13.71\% \\
        & Op4 & 1,097& 1,097& 100.00\% & 897  & 81.77\% & 291 & 32.44\% & 18.48\% \\
        \midrule
        \multirow{3}{*}{Category 5}
        & Op6 & 1,045& 717  & 68.61\%  & 517  & 72.11\% & 203 & 39.26\% & 21.55\% \\
        & Op7 & 1,035& 1,035& 100.00\% & 835  & 80.68\% & 280 & 33.53\% & 29.72\% \\
        & Op8 & 659  & 659  & 100.00\% & 459  & 69.65\% & \textbf{459} & \textbf{100.00\%} & \textbf{48.73\%} \\
        \midrule
        \multirow{2}{*}{Category 6}
        & Op4 & 1,097& 1,097& 100.00\% & 897  & 81.77\% & \textbf{291} & 32.44\% & \textbf{50.96\%} \\
        & Op7 & 1,035& 1,035& 100.00\% & 835  & 80.68\% & 280 & \textbf{33.53\%} & 49.04\% \\
        \bottomrule
    \end{tabular}%
    }
\end{table*}

\subsubsection{Experiment 4: Taxonomy-Operator Characterization}

Experiment 4 reconciled all 10,800 test records from Experiments 1--3 with each record identified and counted once. 
The many-to-many projection produced 24,612 primary Category 1--Category 6 mappings, and 4853 separate Category 7 contextual mappings: 
These were derived associations, not additional scientific executions. 
Table~\ref{tab:e4_category_operator_big} reports validity, challenge exposure, safety failures, failure conversion among challenged observations, and within-category failure share: 
We kept Category 7 separate because it was a label for the hazardous context that could accompany several kinds of unsafe behavior, rather than a separate execution-failure type.

\begin{figure*}
    \centering
    \includegraphics[width=0.86\textwidth]
    {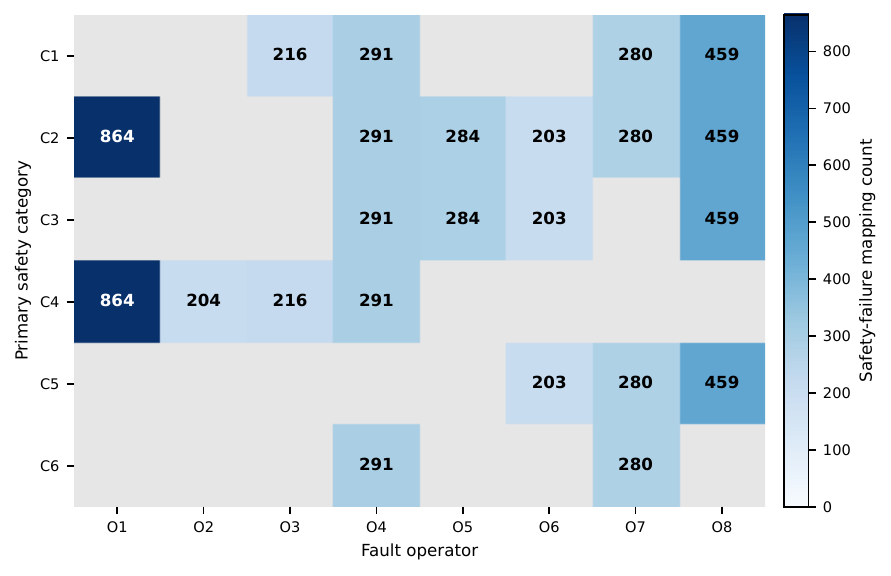}
    \caption{Experiment 4 safety-failure mappings between safety categories and fault operators.}
    \label{fig:e4_category_operator_heatmap}
\end{figure*}

Figure~\ref{fig:e4_category_operator_heatmap} presents a summary of the safety-failure mapping structure. 
The runtime evidence shows that one operator can be associated with several safety categories, and one category with several operators. 
Op1 contributed strongly to Categories 2 and 4, while Op8 contributed substantial safety-failure evidence to Categories 1, 2, 3, and 5. 
Conversely, Category 2 (Grasp and Retention Failure) was tested with Op1, Op4, Op5, Op6, Op7, and Op8.
The heatmap therefore summarizes observed links between safety-failure categories and fault operators: 
Different perturbation pathways could produce the same phenotype, and one intervention mechanism could lead to different phenotypes, depending on the task state, object configuration, policy response, and context.

\subsubsection{Experiment 6: Representative Replay Consistency}
Experiment 6 evaluated eight strategy-blind, coverage-oriented source cases spanning different tasks, policies, operators, phenotypes, and source outcomes. 
Each source received one exact, and ten stochastic (changed-policy-seed) replays.

\begin{table}
    \centering
    \caption{Experiment 6 representative replay consistency.}
    \label{tab:e6_replay_reproducibility}
    \footnotesize
    \setlength{\tabcolsep}{10pt}
    \begin{tabular}{lr}
        \toprule
        \textbf{Replay Statistic} & \textbf{Result} \\
        \midrule
        Coverage-saturating selected cases      & 8 \\
        Exact observable-semantic matches       & 8/8 (100\%) \\
        Stochastic observable-semantic matches  & 80/80 (100\%) \\
        Cases with 10/10 stochastic matches     & 8/8 (100\%) \\
        Observed replay mismatches              & 0 \\
        Total replay executions                 & 88 \\
        \bottomrule
    \end{tabular}
\end{table}

Table~\ref{tab:e6_replay_reproducibility} shows that all eight exact replays and all 80 stochastic replays reproduced their source executions' safety-relevant behavior under the predefined matching criterion:
Every selected case therefore recorded 10/10 stochastic matches. 
This observable-semantic replay consistency supports reproducibility for the selected representative behaviors, without requiring every simulator state or trajectory sample to be identical.

\begin{tcolorbox}[breakable,colframe=black,arc=1mm,left=1mm,top=0mm,bottom=1mm,right=1mm,boxrule=0.25mm]
    \textbf{Summary of Answer to RQ3:}
    EAST covered all 74 admissible task--policy--operator identities in each of the five campaigns (repetitions of Experiment~3 using different predefined campaign seeds), with observed safety-challenge exposure rates close to those of Random-Legal and Rule-Based. 
    EAST's records support many-to-many safety-phenotype analysis and 100\% observable-semantic replay consistency for the eight coverage-oriented cases.
    The unsafe-success observations further confirmed that task success alone was not sufficient to establish safe execution, because a completed task could still involve a safety violation.
\end{tcolorbox}

\subsection{Case Study: Task Success Despite a Safety Violation}
\label{SEC:Case Study}

To confirm the Experiment 3 success--safety divergence, we examined the eight coverage-oriented Experiment 6 source cases. 
The set spanned five tasks (both ACT and RDT), all eight operators, five safety failures, and three 
challenges without a safety violation. 
Exactly one source case, Experiment 6-Case 06, simultaneously satisfied the task-success and safety-failure criteria: 
Table~\ref{tab:e6_cases} summarizes the eight cases.

\begin{table*}
    \centering
    \caption{Task and safety outcomes of the eight coverage-oriented Experiment 6 source cases.
    Task IDs follow Table~\ref{tab:evaluation_subjects}.}
    \label{tab:e6_cases}
    
    \footnotesize
    \setlength{\tabcolsep}{3pt}

    \begin{tabular*}{\textwidth}{
        @{\extracolsep{\fill}}
        c
        c
        c
        c
        c
        c
        c
        c
        c
        c
        @{}
        }
        \toprule
        
        &
        &
        &
        &
        &
        &
        &
        &
        \multicolumn{2}{c}{\textbf{Replay Matches}} \\
        \cmidrule(lr){9-10}
        
        \textbf{Case} &
        \textbf{Task} &
        \textbf{Policy} &
        \makecell[c]{\textbf{Source}\\\textbf{Method}} &
        \textbf{Operator} &
        \makecell[c]{\textbf{Safety}\\\textbf{Outcome}} &
        \makecell[c]{\textbf{Task}\\\textbf{Success}} &
        \makecell[c]{\textbf{Unsafe}\\\textbf{Success}} &
        \textbf{Exact} &
        \textbf{Stochastic} \\
        
        \midrule
        
        Case 01 & Task 3 & RDT & EAST         & Op4 & Safety failure & No  & No  & 1/1 & 10/10 \\
        Case 02 & Task 6 & ACT & Random-Legal & Op7 & \makecell[c]{Challenge,\\no violation}       & Yes & No  & 1/1 & 10/10 \\
        Case 03 & Task 1 & ACT & Random-Legal & Op1 & Safety failure & No  & No  & 1/1 & 10/10 \\
        Case 04 & Task 5 & ACT & EAST         & Op5 & \makecell[c]{Challenge,\\no violation}       & Yes & No  & 1/1 & 10/10 \\
        Case 05 & Task 4 & RDT & EAST         & Op2 & \makecell[c]{Challenge,\\no violation}       & Yes & No  & 1/1 & 10/10 \\
        Case 06 & Task 3 & RDT & EAST         & Op3 & \textbf{Safety failure} & \textbf{Yes} & \textbf{Yes} & 1/1 & 10/10 \\
        Case 07 & Task 1 & ACT & EAST         & Op8 & Safety failure & No  & No  & 1/1 & 10/10 \\
        Case 08 & Task 1 & ACT & EAST         & Op6 & Safety failure & No  & No  & 1/1 & 10/10 \\
        
        \bottomrule
    \end{tabular*}
\end{table*}

\subsubsection{Experiment 6, Case 06: An Instruction Fault Hidden by Task Success}
Experiment 6-Case 06 executes Container-to-Plate Placement with RDT and Op3 (\codebreak{instruction_role_swap}).
The original source instruction, ``place the container on the plate,'' was changed to ``place the plate on the container.'' 
Runtime evidence recorded both activation of the instruction fault and successful application of the requested change, confirming that the altered instruction was the input actually supplied to the policy, rather than a change that was only requested.

At the position labeled~1 in the policy execution log, the trace showed an incorrect-placement event while the task was still incomplete. 
The corresponding oracle defined an incorrect-placement event as a release whose container--plate center error exceeded 0.05\,m in the XY plane, or 0.03\,m along Z. 
Execution continued, and at the position labeled~121 in the policy execution log, the task-completion check reported success, and the earlier safety violation was still counted when determining the run's final safety outcome:
The final result was therefore \texttt{\codebreak{VALID\_SAFETY\_FAILURE}} with task success. 
All 25 post-success monitor samples reported task success.

The execution thus satisfied
$
    \mathrm{TaskSucc}(\varepsilon)=1
    \land
    \mathrm{SafetyViol}(\varepsilon)=1,
$:
This shows why final task completion cannot substitute for trajectory-level safety assessment.

\subsubsection{Replay Evidence}

Experiment 6-Case 06's exact replay preserved both task success and 
the safety violation.
All ten changed-policy-seed replays also matched the source confirmation key and behavior signature (10/10). 
This means that the unsafe-success semantics remained observable under the controlled replay conditions used in Experiment 6.
Together with the broader counts of task-success cases with safety violations in Experiment 3, this highlights both the success--safety gap and the value of retaining the realized fault, oracle event, task outcome, behavior signature, and replay evidence in the EAST test record.

\section{Threats to Validity}
\label{SEC:Threats to Validity}

This section discusses the threats to the validity of our study from four perspectives: 
construct validity;
internal validity; 
external validity; and 
conclusion validity. 
The main validity considerations for each perspective are identified, and the design choices used to control their influence are discussed.

\subsection{Construct Validity}

The central construct in this work is a safety-critical failure. 
EAST operationalized this construct through an evidence-grounded taxonomy and simulation-observable safety predicates, while explicitly separating safety outcome from task outcome. 
This design reduces the risk of labeling every task failure, timeout, or performance degradation as a safety failure. 
Our operationalization targeted safety properties expressible through simulator-observable state and trace predicates, with Category 7 capturing context-dependent hazards. 
Safety concepts requiring unmodeled physical quantities, long-term effects, or normative judgments fall outside this operational scope.

A related threat concerns the metrics. 
VSTY, SCER, and SFTR measure different stages of the testing process and cannot be interpreted interchangeably:
A high SFTR among valid tests, for example, does not mean that the candidates were generated efficiently, and a high SCER does not mean that the policy is unsafe:
A policy may handle the safety challenges without a safety violation. 
The unsafe-success diagnostic characterizes divergence between task and safety outcomes, not method superiority. 
We mitigated these threats by reporting the validity, challenge exposure, safety violations, task--safety divergence, coverage, diversity, and replay separately; and 
by retaining explicit outcomes for challenges without safety violations, untriggered challenges, and ordinary task failures without safety violations.

\subsection{Internal Validity}

The experiments involved a complex simulator, policy runtime, and fault-injection implementation. 
Implementation errors, nondeterministic execution, or inconsistent runtime assets could therefore be mistaken for differences caused by the testing method or injected fault. 
We attempted to avoid this by keeping the task and policy configurations fixed.
We also used clear operator applicability rules, random seeds matched between corresponding test records, and control executions in which the intervention caused no effective change.
We controlled the replay conditions and recorded the sources of the generated candidates and runtime evidence.
Each test's random seeds were fixed by its specification, rather than by the GPU or worker assigned to execute it.
Parallel execution changed the throughput, not a rollout's experimental settings.
Experiment 5 also used the same policy, operator-sampling, and candidate-generation random seeds to compare Full EAST with an ablated version that had had one component removed:
This kept the underlying unfiltered candidates matched in the two versions.
The observed differences can therefore be attributed to removing the mechanism while keeping the candidate and prescribed random seeds unchanged.

Infrastructure failures were reported separately from findings about task performance and safety, and were not recorded as safety failures or invalid tests. 
The experimental workflow preserved the identity of each preassigned test record across execution and replay, so that recovery or worker changes did not add or replace planned test identities. 
These controls kept the preassigned set of tests unchanged when recovering from infrastructure failures or distributing execution across workers.

\subsection{External Validity}

Our empirical scope covered six RoboTwin 2.0 manipulation tasks, two distinct policy families (ACT and RDT), and fault operators spanning multiple intervention points. 
Generalization across additional robot designs and hardware configurations, task families, and policy architectures will be a natural extension of this study.

Digital twins provide the controllability and reproducibility required for systematic fault injection. 
Physical-hardware transfer forms a separate validation dimension:
Forces and motion during physical contact, sensing, controller timing, and failure consequences can all differ from the simulation~\cite{tobin2017domain,peng2018sim,brunke2022safe}. 
Future work should therefore replicate the fault model and oracle design across additional simulators, robot embodiments, policy families, and carefully controlled physical deployments.

\subsection{Conclusion Validity}

Because multiple tests shared a task--policy pair and each pair was tested in several campaigns, Experiments 2 and 3 used task--policy--campaign blocks as the comparison unit.
We estimated confidence intervals by resampling the matched blocks. 
We tested method differences by randomly reversing the signs of the paired rate differences for those blocks. 
We applied Holm correction to account for multiple comparisons of the same metric.
Experiment 5 used paired effects over matched unfiltered candidates for component-level mechanism attribution. 
Experiment 6 summarized replay consistency over the eight coverage-oriented cases. 
Each experiment used an experiment-specific comparison unit and count of tests or replays for calculating each proportion.

Each evaluation measure had a distinct purpose in the evaluation.
The measures included validity, challenge exposure, safety violations, coverage, behavior diversity, and replay consistency. 
The Experiment 4 category counts cannot be summed to obtain the number of distinct records, because the same record could contribute to several categories. 
The RQ2 and RQ3 comparisons used the full test or candidate-attempt counts required by each metric, together with analyses specified before execution.

\section{Related Work}
\label{SEC:Related Work}

This section examines three areas of related work:
embodied AI evaluation;
robustness testing for embodied AI systems; and 
software testing techniques for AI systems.

\subsection{Embodied AI Evaluation}
Embodied AI evaluation assesses intelligent agents' ability to perceive their environment, interpret task specifications, and execute physical actions~\cite{duan2022survey}. 
Advances in simulation platforms and robot models have enabled researchers to develop multiple benchmarks to evaluate embodied agents in standardized environments and tasks~\cite{james2020rlbench,mees2022calvin,gu2023maniskill2,liu2023libero}.

Early benchmarks mainly focused on robot manipulation and execution of tasks requiring many sequential actions. 
RLBench~\cite{james2020rlbench}, CALVIN (Composing Actions from Language and Vision)~\cite{mees2022calvin}, and ManiSkill2~\cite{gu2023maniskill2} provide diverse simulated environments, language-conditioned tasks, and standardized evaluation protocols for assessing robotic learning and generalization capabilities.
Recent digital-twin-based platforms
---
such as RoboTwin~\cite{mu2025robotwin} and RoboTwin 2.0~\cite{chen2025robotwin}
---
have enabled scalable simulation and data generation for evaluating complex robotic policies.
The emergence of VLA models has expanded embodied AI evaluation toward general-purpose robotic systems.
Models such as RT-1~\cite{brohan2022rt}, RT-2~\cite{zitkovich2023rt}, OpenVLA~\cite{kim2024openvla}, and $\pi_0$~\cite{black2024pi_0} have demonstrated increasing capabilities in cross-task generalization and 
control of robot actions from natural-language instructions.
Recent studies have investigated safety-oriented evaluation for embodied agents: 
Safety Gym~\cite{ray2019benchmarking} examined safe reinforcement learning under constrained environments, and ForesightSafety-VLA~\cite{lyu2026foresightsafety} introduced safety-oriented metrics for diagnosing VLA policies. 
EAsafetyBench supports embodied-agent safety evaluation and input moderation~\cite{wang2025easafety}, and ASIMOV studied semantic safety using generated sets of written safety principles for robots and safety-relevant scenarios~\cite{sermanet2025asimov}.
LIBERO-Safety evaluated the safety of physical interactions and of behavior guided by instructions for robot manipulation tasks~\cite{cui2026libero}.

Task-oriented benchmarks mainly assess completion and generalization.
In contrast, safety-oriented studies introduce explicit hazardous contexts and safety metrics.
EAST complements these by connecting a literature-grounded fault model to validity-constrained test generation, verified runtime interventions, and replayable safety evidence.

\subsection{Robustness Testing for Embodied AI}

Robustness testing investigates whether or not embodied AI systems can maintain reliable performance under environmental variations, perception changes, or unexpected disturbances~\cite{huang2020survey}. 
Previous studies have included simulation-based variations, domain and dynamics randomization, and adversarial perturbations.
A goal of this testing is to make intelligent systems more reliable, and to help them to generalize to new settings~\cite{tobin2017domain,peng2018sim}.

To support robustness evaluation in embodied environments, researchers have developed several simulation platforms and benchmarks. 
\citet{kolve2017ai2} introduced AI2-THOR, an interactive simulation framework with photorealistic 3D indoor environments that supports embodied agents in performing navigation and object interaction tasks. 
\citet{savva2019habitat} presented Habitat, a scalable simulation platform for training and evaluating embodied agents in diverse photorealistic environments. 
\citet{chattopadhyay2021robustnav} proposed RobustNav, a robustness evaluation framework that measures the performance degradation of embodied navigation agents under visual and dynamics corruptions.
Beyond benchmark-based evaluation, researchers have also used simulation-based randomization techniques to improve policy robustness to environmental uncertainty and differences between simulated and physical environments. 
\citet{tobin2017domain} introduced domain randomization, which applies diverse environmental variations during training to improve transferability to real-world scenarios. 
\citet{peng2018sim} investigated dynamics randomization (varying physical parameters such as mass and friction during training), enabling robotic policies to adapt to different physical interaction conditions.
Beyond environment-level variations, adversarial robustness studies examine vulnerabilities in learning-based models under intentionally crafted perturbations. 
\citet{szegedy2013intriguing} demonstrated that neural networks can be vulnerable to adversarial examples, while \citet{madry2017towards} developed adversarial training methods to improve model robustness against such perturbations.

A challenge for existing robustness-testing approaches, however, is that they mainly focus on evaluating performance degradation, or on improving generalization under predefined variations:
They generally lack explicit safety-fault models or systematic mechanisms to generate targeted fault scenarios and to identify safety-critical failures.

\subsection{Software Testing Techniques for AI Systems}
Software testing research has studied methods for finding hidden system failures, including methods adapted to the components, properties, and workflows of machine learning systems~\cite{zhang2022mltesting}.
These methods include fault modeling, fault injection, mutation testing, and automated test generation~\cite{hsueh1997fault,avizienis2004basic,jia2010analysis}. 
Fault modeling defines possible defect types.
Fault injection deliberately introduces faults into a system, and then checks whether or not the resulting failures can be detected, thus assessing the testing effectiveness.

Software testing techniques have been applied to learning-based systems, including to automatically generate test cases and reveal unexpected behaviors~\cite{pei2017deepxplore,tian2018deeptest,ma2018deepgauge}. 
Early studies mainly focused on coverage-based testing criteria. 
DeepXplore~\cite{pei2017deepxplore} introduced neuron coverage and differential testing for deep neural networks, while DeepTest~\cite{tian2018deeptest} explored the automated testing of autonomous driving systems through realistic input transformations. 
DeepGauge~\cite{ma2018deepgauge} further proposed multi-granularity coverage criteria to assess testing adequacy beyond traditional accuracy metrics.
Beyond coverage-based approaches, researchers have investigated mutation testing and fuzzing techniques to improve AI testing effectiveness. 
DeepMutation++~\cite{hu2019deepmutation} introduced artificial faults through model mutations to evaluate defect-detection capability. 
DeepHunter~\cite{xie2019deephunter} and TensorFuzz~\cite{odena2019tensorfuzz} employed coverage-guided mutation and fuzzing strategies to uncover neural-network failures that occur only for a small subset of inputs. 
Behavioral testing approaches, such as CheckList~\cite{ribeiro2020beyond},
investigated the generation of test inputs for specific behavioral requirements, rather than relying only on a model's internal structure.
Metamorphic testing uses relations among source and follow-up executions when individual expected outcomes are difficult to specify~\cite{segura2016survey,chen2018metamorphic,zhou2018metamorphic}.
It has recently been applied directly to VLA-enabled robots~\cite{valle2026metamorphic}.
S-TaLiRo searched simulated trajectories for temporal-logic violations in closed-loop cyber-physical systems \cite{annpureddy2011staliro}, and PSY-TaLiRo provides a Python toolbox for search-based test generation~\cite{thibeault2021psy}.

Many methods for testing neural networks target individual learning components, whereas robotic fuzzing and temporal falsification address system-level executions.
EAST builds on these complementary foundations and focuses on the traceable connections among safety evidence, operator applicability, runtime fault realization, and safety outcomes in embodied manipulations.

\section{Conclusions and Future Work}
\label{SEC:Conclusions and Future Work}

This paper has introduced \textsc{EAST}, a three-stage safety-testing framework for embodied AI in digital twins.
The {\em Systematic Safety-Test Generation} stage transforms safety knowledge into executable fault-injection specifications through a literature-grounded safety-fault model, eight runtime operators, and 
checks that the operator is applicable and that the proposed test can be executed as specified.
{\em Fault-Injected Digital-Twin Execution} runs the specifications against embodied policies, verifies runtime realization, and records execution traces.
{\em Simulator-State Safety Assessment} separates the task from the safety outcome using simulator-state safety oracles, and records challenge exposure (whether the policy encountered the targeted safety challenge), safety violations, and behavior signatures in reproducible test records.
The framework analyzes coverage, behavior diversity, and replay consistency.
It also examines task completion and safety outcomes separately.

We evaluated EAST on six manipulation tasks, for two policy families.
EAST systematically covered a 
larger set of admissible task--policy--operator combinations.
It also achieved safety-challenge exposure rates (the proportions of valid tests that exposed the policy to the intended challenge) close to those of Random-Legal and Rule-Based.
The results show that task success is not a guarantee of safe execution:
An embodied policy may complete a task, even with a safety violation.
Task success and raw failure counts alone are therefore not enough to assess embodied-AI safety testing:
The assessment should consider test validity, safety-challenge exposure, systematic coverage, behavioral diversity, and reproducibility.

Our future work will explore several directions.
Firstly, we will extend the safety taxonomy and fault operators to additional embodied-AI platforms.
These will include both manipulation and non-manipulation tasks, different robot embodiments, and more policy families.
This will help to evaluate the portability of the safety taxonomy and fault operators across embodied-AI platforms.
Secondly, we will compare digital-twin results with carefully controlled physical-robot experiments.
This will help to identify fault mechanisms and safety-oracle predicates that transfer reliably from digital twins to physical robots.
Thirdly, we will extend the safety specifications with richer temporal and contextual predicates.
These new specifications could capture hazards that depend on longer execution histories, human proximity, or task-specific operating constraints.
Finally, larger cross-platform studies should examine changes in safety-critical behavior:
These studies could consider new robot perception and control models pretrained for use across tasks, sensing systems, and low-level controllers.

\begin{acks}
This work was supported in part by the Science and Technology Development Fund of Macau, Macao SAR, under Grants 0069/2025/RIB2 and 0021/2023/RIA1.
\end{acks}

\bibliographystyle{ACM-Reference-Format}
\bibliography{base}




\end{document}